\documentclass[a4paper,10pt,twoside]{article}
\usepackage{amsmath} 
\usepackage{amssymb} 
\usepackage{amsthm}  

\usepackage{fancyhdr}
\usepackage{enumitem}
\usepackage[utf8]{inputenc}
\usepackage{setspace}
\usepackage{graphicx}
\usepackage{textcomp}
\usepackage[retainorgcmds]{IEEEtrantools}
\usepackage{url}

\usepackage[a4paper,twoside, hmarginratio={1:1}]{geometry}

\newcommand{\1}{\frac{1}{2}}
\newcommand{\oo}[1]{\overline{#1}}

\def\tb{t_\beta}
\def\LL{\mathcal{L}}
\def\OO{\mathcal{O}}
\newcommand{\grandO}[1]{O\mathopen{}\left(#1\right)}
\def\ma{M_{A_0}}
\def\dslash{d\!\!\!/}

\def\lhep{\texttt{LanHEP} }
\def\PD{P_{\partial}}
\def\Mathematica{\texttt{Mathematica} }

\begin{document}
\begin{titlepage}
\begin{center}

\vspace*{3cm}

{\Large {\bf SUSY Higgs searches : beyond the MSSM}}

\vspace{8mm}

{F. Boudjema${}^{1)}$, G. Drieu La Rochelle${}^{1)}$}\\

\vspace{4mm}

{\it 1) LAPTh$^\dagger$, Univ. de Savoie, CNRS, B.P.110,
Annecy-le-Vieux F-74941, France}

\vspace{10mm}

\today
\end{center}

\centerline{ {\bf Abstract} } \baselineskip=14pt \noindent

{\small The recent results from the ATLAS and CMS collaborations
show that the allowed range for a Standard Model Higgs boson is
now restricted to a very thin region, if one excludes a very heavy Higgs. Although those limits are
presented exclusively in the framework of the SM, the searches
themselves remain sensitive to other Higgs models. We recast the
limits within a generic supersymmetric framework that goes beyond
the usual minimal extension. Such a generic model can be
parameterised through a supersymmetric effective Lagrangian with
higher order operators appearing in the K\"ahler potential and the
superpotential, an approach whose first motivation is to alleviate
the fine-tuning problem in supersymmetry with the most dramatic
consequence being a substantial increase in the mass of the
lightest Higgs boson as compared to the minimal supersymmetic
model. We investigate in this paper the constraints set by the LHC
on such models. We also investigate how the
present picture will change when gathering more luminosity. Issues
of how to combine and exploit data from the LHC  dedicated to searches
for the standard model Higgs to such supersymmetry inspired
scenarios are discussed. We also discuss the impact of invisible decays of the Higgs in such scenarios.}

\vspace*{\fill}

\vspace*{0.1cm} \rightline{LAPTh-046/11}

\vspace*{1cm}

$^\dagger${\small UMR 5108 du CNRS, associ\'ee  \`a l'Universit\'e
de Savoie.} \normalsize

\vspace*{2cm}

\end{titlepage}

\renewcommand{\topfraction}{0.85}
\renewcommand{\textfraction}{0.1}
\renewcommand{\floatpagefraction}{0.75}

\section{Introduction}
The impressive 2011 collection of data from the Large Hadron
Collider (LHC) provides us with an outstanding insight on many
different theoretical models, ranging from the Standard Model to
more complex and exotic constructions. Among the latter,
supersymmetry is certainly not the least scrutinized. So far no
evidence has been found for any new signal in supersymmetry. This
leads one to ponder on two complementary issues. On the one hand,
in the case of the search for superpartners, the scale  of
supersymmetry breaking can now be seen to be driven to the TeV
range which from the point of view of naturalness puts the
original motivation of supersymmetry in jeopardy, even if strictly
speaking the argument applies to the third family and the higgsino
sector which are directly connected to the Higgs. On the other
hand, the current limit on the Standard Model, SM,  Higgs
cornering its mass in the narrow range between 115 and 140 GeV,
where the experimental sensitivity has not enough exclusion power
yet and will not probably do so until 2012, is very much
compatible with the most simple supersymmetric implementation of
the Higgs. Nonetheless since supersymmetry does not seem to be
``around the corner" leading to a stronger and stronger tension
with the naturalness argument, it has been
suggested
\cite{brignole}, \cite{dine_bmssmhiggs_0707}, \cite{antoniadis_bmssm_0708,antoniadis_bmssmhiggs_0806,antoniadis_bmssmhiggs_0910,antoniadis_bmssmhiggs_1012},
\cite{ponton_susy_higgs_0809}, \cite{carena_bmssmhiggs_0909,carena_bmssmhiggs_1005,carena_bmssmhiggs_1111}
that new heavy  supersymmetric physics can contribute to higher
order operators that improve fine-tuning\cite{espinosa_finetuning_2004,ross_finetuning_0903,cassel_bmssmhiggs_1103}. Another consequence is
the possibility, through these new operators, for the lightest
supersymmetric Higgs to be much heavier than the usual lightest
Higgs of the MSSM (the Minimal Supersymmetric Standard Model).
Since the properties and couplings of the Higgses in such BMSSM
(Beyond the Minimal Supersymmetric Standard Model) scenarios can
be very much modified, it is important to inquire whether such
Higgses may be hiding in a wider range than the present narrow
range that applies for the standard model Higgs.\\
\noindent A generic analysis for supersymmetric Higgses does not exist
since the supersymmetric models are numerous and often lead to
different phenomenologies. In addition, the Standard Model Higgs
analyses are not straightforward to interpret in a supersymmetric
framework, where productions and branching ratios can be greatly
enhanced or suppressed. This explains perhaps why the question of
excluding the existence of supersymmetric Higgses (and conversely,
looking for them if they do not show up in standard model type
searches) needs a dedicated study. A common misbelief is to think
that because the mass of the lightest Higgs is between 114 GeV and
135-140 GeV, only a few decay modes are allowed. This may be true
in the simplest realisation of supersymmetry, the MSSM, but is no
longer true for other models, for instance the Next-to Minimal
Supersymmetric Standard Model (NMSSM) allows for a lightest Higgs
which is much heavier (ref \cite{ellwanger_nmssm}). The 135 GeV
upper bound on the lightest Higgs is a feature of the MSSM, not
supersymmetry itself. A similar concern is that the lightest Higgs
is not necessarily so SM-like : for instance its branching ratio
to invisible particles can be important. Thus, hunting the Higgs
bosons in the supersymmetric landscape is much of an exploit.\\
Nevertheless, workarounds do exist. Indeed, most of the difficulty
comes from the fact that, if we already know that the MSSM
spectrum has to be part of any supersymmetric theory, there are no
constraints for additional particles, which means the coexistence
of many models. However, if one is only interested in the low
energy effects of those particles, one can pursue a model
independent approach through an effective field theory, EFT, (see
\cite{burgess_eft} for a review of EFT, among others). This will
be the context of the present study : we assume a decoupling
between a low energy theory -- the MSSM -- and new particles, so
that the physical effects of the latter will be accounted for by
higher-dimensional operators in the supersymmetric potentials. It
is well-known (see \cite{brignole}, \cite{dine_bmssmhiggs_0707}, \cite{antoniadis_bmssm_0708,antoniadis_bmssmhiggs_0806,antoniadis_bmssmhiggs_0910,antoniadis_bmssmhiggs_1012},
\cite{ponton_susy_higgs_0809}, \cite{carena_bmssmhiggs_0909,carena_bmssmhiggs_1005,carena_bmssmhiggs_1111})
that this approach reproduces a wide range of the SUSY Higgs
phenomenology. In fact we do not even require a full decoupling
between the MSSM spectrum and the new particles, but only a
decoupling between the typical energy scale of the Higgs processes
at the LHC and the new particles. This approach has the immediate
drawback of not accounting for light degrees of freedom, as may
exist in the NMSSM with a light singlet for instance : hence some
of the BMSSM scenarios will be missed. However, those models
require a different study, whereas the effective approach can use
the usual MSSM decay channels for Higgses.\\
While completing this paper an analysis within the same model
taking into account LHC data appeared. As we will show our
analysis covers ref \cite{carena_bmssmhiggs_1111} and goes beyond
in considering a few key issues having to do with how to exploit
LHC data made in the context of the Standard Model Higgs to models
that do not necessarily share the properties of the Standard
Model. With the latest ATLAS and CMS combined data (see
\cite{atlas_cms_lp_comb}), we comment on the uncertainties one
introduces by attempting to combine data outside the
collaborations. In our analysis we cover a larger scan in $\tan
\beta$. Moreover, although for most of our study the
supersymmetric particles do not play a role, we dedicate a small
section to the impact of light enough neutralinos that could
contribute to the invisible width of the Higgs. On the theoretical
side, we show how one can exploit and extend automated tools such
as {\tt lanHEP}\cite{lanhep1,lanhep2} to derive from the
superfield formulation the effective Lagrangian in component
fields that
could easily be fed into matrix element generators.\\

The article will be organised as follow : in Section 2, we
describe the new operators arising in the context of the EFT and
their implementation in the aim of studying the Higgs
phenomenology. In Section 3 we explain how the Feynman rules of
the model are computed which leads to Section 4 which describes
how the computation of the relevant observables for Higgs physics
is performed. Section 5 is the analysis itself which exploits data
from the LHC to derive the latest constraints on the model and
will give prospects with the next round of accumulated data.
Section~6 outlines our key findings and conclusions. Some
technical details of our calculation covering the effective
Lagrangian expressed in field components is left to an Appendix.

As shorthand notation, we write $c_\theta$ for $\cos \theta$ and similarly for other trigonometric functions.

\section{Effective Field Theory}
As in any effective field theory, one has to start by defining the
low energy theory, and the scale of the heavy spectrum (see
\cite{burgess_eft}). We will do so by fixing the leading order lagrangian to be the one of the MSSM :
\begin{equation}
\LL^{(0)}=\LL_{MSSM}
\end{equation}
and the scale of new physics to be
\begin{equation}
M\sim1.5\textrm{ TeV}.
\end{equation}
The choice of the scale is fixed
in order to have a valid effective expansion for relevant
observables in the physics under study. Reference\cite{wudka_piriz_bmssm_1997} catalogues a large number of
operators up to dimension-6  within a supersymmetric framework. Here, we  are only  interested in Higgs
phenomenology at colliders where the effect of the high dimension operators is most drastic. A scale above one TeV seems a
reasonable choice for Higgs studies even for for LHC energies. Note that the superpartner spectrum can be
lower or higher than the scale $M$, without affecting the validity
of the effective expansion. The effective expansion is obtained by
integrating out the extra heavy particles : the resulting terms
are going to enter any of the supersymmetric potentials, namely
the K\"ahler potential $K$, the superpotential $W$, and the
susy-breaking potential. Developing the former two, we get :
\begin{align} K_{\textrm{eff}}&=
K^{(0)}+\frac{1}{M}K^{(1)}+\frac{1}{M^2}K^{(2)}+\grandO{\frac{1}{M^3}}\label{effK}\\
W_{\textrm{eff}}&=
W^{(0)}+\frac{1}{M}W^{(1)}+\frac{1}{M^2}W^{(2)}+\grandO{\frac{1}{M^3}}
\end{align}
The choice of truncating the effective theory at a
given order -- here at order 2 -- is arbitrary and dictated by the
required accuracy of the predictions and the computability of the
higher orders at the same time. References
\cite{carena_bmssmhiggs_0909} and
\cite{antoniadis_bmssmhiggs_0910} have shown independently that
the $\grandO{1/M^2}$ corrections to the Higgs masses are still of the
order of 10 GeV, hence we decided to take them into account, but
going to third order implied a lot of complications for relatively
small changes. For simplicity, and since it should retain most of
the contribution to the Higgs observables, we will restrict
ourselves to operators involving solely Higgs superfields. Then
there is only a handful of those operators in dimension 5 and 6, as was previously
shown in \cite{brignole,carena_bmssmhiggs_0909,antoniadis_bmssmhiggs_0910}.
\begin{IEEEeqnarray}{rCl}
W_{\text{eff}}&=&\zeta_1\frac{1}{M}\left(H_1.H_2\right)^2\\
K_{\text{eff}}&=&a_1\frac{1}{M^2}\left(H_1^{\dag}e^{V_1}H_1\right)^2+a_2\frac{1}{M^2}\left(H_2^{\dag}e^{V_2}H_2\right)^2+a_3\frac{1}{M^2}\left(H_1^{\dag}e^{V_1}H_1\right)\left(H_2^{\dag}e^{V_2}H_2\right)\nonumber\\
&&+a_4\frac{1}{M^2}\bigl(H_1.H_2\bigr)\left(H_1^{\dag}.H_2^{\dag}\right)+\:\frac{1}{M^2}\left(H_1.H_2+H_1^{\dag}.H_2^{\dag}\right)\left(a_5H_1^{\dag}e^{V_1}H_1+a_6H_2^{\dag}e^{V_2}H_2\right)
\end{IEEEeqnarray} where $H_1,H_2$ are the two Higgs superfields in
the gauge basis, assuming the hypercharges to be $Y_1=-1,Y_2=1$. $V_1,V_2$ are the gauge superfields in the representation of $H_1$ and $H_2$ respectively. Note that we introduced the new parameters $a_i,\
\zeta_1$, that will be referred to as effective coefficients.
Those are the remnants of the couplings of the heavy particles to
the MSSM spectrum, we will furthermore assume this extra physics
to be weakly coupled, ensuring that those coefficients are at most
of order one.\\ So far we have not dealt with the susy-breaking
potential, and so what we have are the operators generated by an
extra physics that is supersymmetry-conserving. However there is
no such a requirement, since susy breaking can occur above or
below the scale of extra physics. One natural way to extend our
operators to susy-breaking ones is to convert the effective
coefficients into spurions (see \cite{antoniadis_bmssmhiggs_0910}) : \begin{align}
\zeta_1&\longrightarrow\zeta_{10}+\zeta_{11}m_s\theta^2\\
a_i&\longrightarrow
a_{i0}+a_{i1}m_s\theta^2+a_{i1}^*m_s\overline{\theta}^2+a_{i2}m_s^2\overline{\theta}^2\theta^2
\end{align} We have introduced the spurion breaking mass $m_s$ (which appears with the Grassman basis vector $\theta$),
which for the sake of the effective expansion to hold must be
small compared to $M$. This redistributes the operators among the
different potentials : $a_{i0}$ operators will be in the K\"ahler
potential, ($\zeta_{10},\ a_{i1}$) in the superpotential, and
($\zeta_{11},\ a_{i2}$) in the soft breaking potential. We take all parameters real. The case of imaginary parameters for the dim-5 operators and its impact
on flavour and CP violation is studied in \cite{altmannshofer_bmssmcp_1107}\\
Foreseeing the physical effects of those operators is rather not
intuitive, partly because the scalar potential comes from $F$ and
$D$ terms which are much less straightforward than in the MSSM
case, and partly because new interactions terms will also appear.
To keep track of the new parameters that are introduced, note that
the two parameters $\zeta_{10,11}$ are from dim-5 operators
whereas the much larger set $\{a_{ij}\}$ stem from dim-6
operators. The lagrangian in terms of fields of a generic
supersymmetric theory is given in the Appendix \ref{full_lag}, and
we will outline its features in the following.

\section{Effective Lagrangian}
Whereas the effective theory seems rather simple in its
superfields formulation, this is no longer true when looking at
the Lagrangian in terms of field components. In comparison, in the
MSSM where we also have K\"ahler potential and superpotential
given as superfield functionals :
\begin{equation}
K=K\left(\Phi_i,\oo{\Phi}_{\oo{j}},e^{V_n}\right)\qquad
W=W(\Phi_i)
\end{equation}
 where $i$ labels the chiral superfields in $\Phi$ and $n$ the
different representations of the gauge superfield $V$, the only
quantities appearing in the field-component Lagrangian are
\begin{equation}
K_{i\oo{j}}=\partial_{\Phi^i}\partial_{\oo{\Phi}^{\oo{i}}}K,\qquad
K_n=\partial_{e^{V_n}}K,\qquad W_i=\partial_{\Phi^i}W,\qquad
W_{ij}=\partial_{\Phi^i}\partial_{\Phi^j}W
\end{equation}
 And, from those
quantities only $W_i,W_{ij}$ are non-trivial since it turns out
that
\begin{equation}
K_{i\oo{j}}=\delta_{i\oo{j}}\qquad K_{in}=\oo{\phi}_{\oo{i}}.
\end{equation}
Beyond the MSSM, not only those simplifications are lost but more
derivatives of K and W get involved, as shown in the Appendix. The
case of  the scalar potential is a good example:
\begin{equation}
V_{\textrm{MSSM}}=-\oo{W_i}W^i-\1|K_n\;gT^n|^2\longrightarrow
V=-(\oo{\textbf{W}}^{\oo{i}}+\1
K_{\overline{i}kl}\psi^k\psi^l)K_{\oo{i}j}(\textbf{W}^j+\1
K_{j\oo{k}\oo{l}}\psi^{\oo{k}}\psi^{\oo{l}})-\1|K_n\;gT^n|^2
\end{equation}
where we have now decomposed the superfields into fields
components, that is $\Phi=\binom{\phi}{\psi}$, and the K\"ahler
potential and superpotential are now evaluated on the scalar
fields : $K=K(\phi_i,\oo{\phi}_j)$ and $W=W(\phi_i)$.\\

At this point, it is fair that, though a human processing of the
Lagrangian is still feasible, it may easily lead to a large amount
of mistakes and should be done again any time one wishes to add
another operator, so on the whole does not seem very attractive.
For this reason, we decided to use and develop \lhep, a code
dedicated to the computation of Feynman rules particularly handy
in the supersymmetric framework (\cite{lanhep1,lanhep2}), to
derive the field-component Lagrangian. Development was needed
since \lhep  did not include higher order derivatives of the
K\"ahler potential and the superpotential. By implementing an
extension including all terms of the Lagrangian, together with the
truncation of the Lagrangian at a given order of the effective
expansion, we had an automated tool to deal with any effective
supersymmetric theory.

\subsection{Scalar potential}
Given the Lagrangian in terms of fields, one can now start to link
the initial components of the superfields to the physical fields,
on which phenomenological consideration will apply. The first step
is to obtain standard kinetic terms for all fields. Indeed, since
for electroweak symmetry breaking to occur the Higgs fields
acquire non-vanishing vacuum expectation values, the effective
expansion will exhibit $K_{i\oo{j}}\neq\delta_{i\oo{j}}$, making
kinetic terms no longer trivial. Fortunately, since our new
operators only involves Higgs superfields, this affects only Higgs
and Higgsinos fields, that we denote by
$H_i=\binom{\tilde{h}_i}{h_i}$. Both set of fields will be
transformed by the same matrix $\PD=\sqrt{(K)_{i\oo{j}}}$
\begin{equation}
\binom{h_1'}{h_2'}=\PD\binom{h_1}{h_2}\qquad\binom{\tilde{h}_1'}{\tilde{h}_2'}=\PD\binom{\tilde{h}_1}{\tilde{h}_2}
\end{equation}
The transformation looks unorthodox at first glance since the two
fields do not share the same quantum numbers, but $\PD$ is non
trivial only when the gauge symmetry is broken, which avoids this
consideration.\\ Since all fields are now normalized (from the
quantum theory point of view), we can carry on to the next usual
step : stabilising the potential. This will fix the values of
$v_1',v_2'$, the vevs of the $h'$ fields (or equivalently $v_1,v_2$
vevs of the $h$ fields). At this point there is another departure
from the MSSM : the minimisation constraints can have more than
one non-trivial solution : this new minimum will usually be
shifted to the heavy scale $M$, whereas the MSSM-like minimum
stays at the weak scale. Issues arising when the new minimum
becomes the global one, or tunneling features in-between minima
have already been discussed in \cite{delaunay}, and we will only
consider MSSM-like minimum. Naming $V$ the Higgs scalar potential,
the minimisation equation simply reads
\begin{equation}
d(V_\circ\PD)\,(h'\rightarrow v')=0\qquad\Leftrightarrow\qquad
dV(h\rightarrow v)=0.
\end{equation}
where $d$ is the derivative along $\phi$,
the vector of Higgs scalar fields. The following step is to find a
basis of mass eigenstates, hence diagonalising the mass matrix. In
the case of Higgses we write
\begin{eqnarray}
M_{h'}&=&d^2(V_\circ\PD)\,(v')\\ &=&\left(\PD\cdot
d^2V\cdot\PD\right)_\circ\PD(v')\\ M_{h'}&=&\left(\PD\cdot
d^2V(v)\cdot\PD\right)
\end{eqnarray}
Since $\PD$ and $d^2V$
are hermitian, $M_{h'}$ will be diagonalised by an orthogonal
matrix $P$. The relation between the initial fields and the
physical fields is then
\begin{equation}
\binom{h_1}{h_2}=\PD^{-1}
P\binom{h}{H}
\end{equation}
 in the case of CP-even Higgses.  The same applies to
CP-odd and charged Higgses, with different $\PD$ in the charged
case (since it is gauge-broken). We will denote the lightest CP
even Higgs by $h$ and the heaviest by $H$.\\

While the expression for $\PD$ is fairly simple and only includes
susy-conserving operator in the K\"ahler potential (the exact
expression is given in Appendix \ref{PD}), the mass matrix
$M_{h'}$ expression is rather cumbersome since all operators are
involved and we do not reproduce it here. A similar operation is
performed on Higgsinos, and the rest of the spectrum can be
diagonalised in the conventional way, leading us to the next
computation step.

\subsection{Extracting initial parameters from observables}
At this stage we have to trade some parameters with experimental
data : namely all standard parameters together with the vacuum
expectations values, that is $$g_1,g_2,v_1,v_2.$$ Those parameters
are fixed by the weak bosons masses, the electromagnetic coupling
and the ratio of the vev $\tb=v_2/v_1$.  The last one being so far
undetermined, but can be extracted from decays of the CP-odd Higgs
boson, such as $A_0\rightarrow\tau^+\tau^-$. The mass terms of the weak bosons include a
contribution from the new operators :
\begin{align}
\label{mwmz}
M_Z^2&=\1
(g_1^2+g_2^2)v^2\left[1+\frac{v^2}{M^2}\left(4a_{10}c_\beta^4+4a_{20}s_\beta^4+2a_{50}c_\beta^3s_\beta+2a_{60}c_\beta
s_\beta^3\right)\right] \nonumber \\ M_W^2&=\1
g_1^2v^2\left[1+\frac{v^2}{M^2}\left(2a_{10}c_\beta^4+2a_{20}s_\beta^4+2a_{30}c_\beta^2s_\beta^2+2a_{50}c_\beta^3s_\beta+2a_{60}c_\beta
s_\beta^3\right)\right]
\end{align}
We write $v^2=v_1^2+v_2^2$. We clearly see that  some contributions, $a_{10}, a_{20}$
and $a_{30}$, lead to
$\frac{M_W^2}{M_Z^2}\neq\frac{g_1^2}{g_1^2+g_2^2}$, they do not
maintain the global $SU(2)$ of the Standard Model Higgs. This
particular combination will therefore be very much constrained
through indirect electroweak precision measurements. \\

In order to keep the handy analytic results derived from \lhep,
those equations are solved through \Mathematica making use of the
effective expansion to perform the extraction linearly.
Minimisation of the potential was obtained from the same token. At
this point we have successfully obtained the relation from initial
parameters and fields to physical observables and fields, so we
are ready for the phenomenological study. From all the parameters
of the MSSM and the effective expansion, we will use the following
subset in the analysis : the mass of the CP-odd Higgs $\ma$ and
the ratio of the vevs $\tb$ from the MSSM, $\zeta_{10},\zeta_{11}$
from the first order contribution and $a_{ij}\ (i=1..6,\ j=0..2)$
from the second order. This adds up to a total of an extra 22
(real) parameters (beside those of the MSSM :
$\ma,\tb,m_{\tilde{f}},M_1,\cdots$). This large number of
parameters will be an important issue in the numerical analysis
and how one scans on this parameter space.

\section{Computing observables}
Having found the physical fields and computed their masses,
extracted the parameters and computed the Lagrangian, we are still
far from the complete phenomenology which involves the actual
values of the branching ratios and the cross-sections, two
quantities not so straightforward to obtain considering that Higgs
phenomenology, specially in SUSY models, is plagued with radiative
corrections. However since \lhep did not only provide us with
analytic expressions, but also formatted output to
phenomenological codes such as \texttt{CalcHEP}\cite{calchep} and
\texttt{HDecay}\cite{hdecay}, we merely have to choose which tool
to use. So we considered each observable on a case by case basis
in order to optimise the precision of the calculation within a
reasonable computing time.\\

\subsection{Combining loop and effective expansions}
It is a well-known fact that SUSY Higgs phenomenology cannot be
separated from loop computations. However, we cannot reach the
state of the art prediction obtained in the SM with event
generators and multi-loop computation tools starting from scratch
with a brand new model. Besides, the accuracy would be unnecessary
since we intend to vary freely our effective coefficients in the
interval $[-1,1]$. Depending on the observable, some
radiative corrections will be computed, as we describe now. First,
let us point that we can write any observable as a double
expansion, the loop expansion, and the effective one. We will
categorise observables as follows :

\begin{itemize}
 \item Decorrelated expansion. We mean by this that effective
 couplings and MSSM/QCD loops do not interfere, allowing thus to
 write
 $\OO=\OO^{(0)}+\delta_{\textrm{loop}}\,\OO+\delta_{\textrm{eff}}\,\OO$,
 and we can take separately the prediction for
 $\delta_{\textrm{eff}}\,\OO$ from a tree-level code and
 $\delta_{\textrm{loop}}\,\OO$ from an MSSM-dedicated code. This
 will be the case of masses for instance :
 $\delta_{\textrm{eff}}\,m$ are taken from the \lhep output
 whereas $\delta_{\textrm{loop}}\,m$ are obtained from a spectrum
 calculator code, namely \texttt{Suspect} \cite{suspect}. The
 computation of masses is done as follows : loop-order masses and
 mixing are computed by the spectrum calculator, and an effective
 potential is reconstructed from those values (see for example
 \cite{boudjema_efflambda}), then the potential coming
 from higher order terms ($\grandO{1/M,1/M^2}$) is added and the
 result is eventually diagonalised. For some observables such as
 decay to neutralino/charginos, $\delta_{\textrm{loop}}\,\OO$ will
 be neglected.
 \item Factorisable expansion. This case arises when
 we can factorise the effective expansion from the loop expansion.
 That is to say that the scale factor between the tree-level
 amplitude of the MSSM and the tree-level amplitude in the
 effective theory is the same as the scale factor between the
 one-loop amplitudes in both theories, and so on up to the full
 loop computation, so that this scale factor can safely be applied
 to the full cross-section. This is equivalent to requiring that
 both theories have the same K-factor, for the given observable.
 This is the case of most of the Higgses decay, for instance the
 partial decay width of the lightest Higgs  into $b$ fermions.
 \begin{equation}
\Gamma_{h\rightarrow\oo{b}b}=R_{g_{h\oo{b}b}\ \textrm{eff}}\times\Gamma_{h\rightarrow\oo{b}b\ \textrm{loop}}^{\text{MSSM}}\label{approx_hbb}
 \end{equation}
 where $R_{g_{h\oo{b}b}\ \textrm{eff}}$ is the ratio of the $h\oo{b}b$ coupling in the effective theory to the MSSM one and $\Gamma_{h\rightarrow\oo{b}b\ \textrm{loop}}^{\text{MSSM}}$ is the MSSM partial width.\footnote{Strictly speaking, it is the MSSM partial width obtained by replacing the MSSM masses with effective masses.}
 \item Nested expansion. This happens when the loop contribution
 cannot be neglected, but cannot be factorised either. This
 is the case for observables such as Higgs decays to photons or gluons.
 It occurs first at the one-loop level. In those cases, since the
 effective rescaling of the coupling $g_{h\oo{b}b}$ and
 $g_{h\oo{t}t}$ are different, the effective scale factor cannot be
 factorised. Then relations such as eq. \ref{approx_hbb} do not
 apply and the computation has to be done with the specific ratio,
 which usually means modifying some MSSM-dedicated codes such as \texttt{HDecay}\cite{hdecay}, as will
 be detailed later.
\end{itemize}

At the end of the day, the observables we need are the masses and the product of the production cross-section by the branching ratio to the final state considered : $\sigma\times BR$. Concerning the masses, they are computed with a reasonable accuracy on both sides of the expansion, and in a reasonable time since the effective shift is an analytic formulae, and the loop shift has to be re-evaluated only when changing MSSM parameters.
\begin{equation}
 m=m_\text{loop}+\delta m_{\text{eff}}
\end{equation}
For cross-sections and branching ratios, the experimental results
are usually given rescaled from the SM prediction, so in the case
of factorisable expansion (with respect to the standard model
case) the loop precision is more than sufficient and the
computation is straightforward since effective ratios are given as
analytic formulas. Examples are decays to fermions or weak gauge
bosons :
\begin{IEEEeqnarray}{cc}
 \frac{BR_{h\rightarrow\tau\tau}}{BR_{h\rightarrow\tau\tau\ SM}}=\left(\frac{g_{h\oo{\tau}\tau}}{g_{h\oo{\tau}\tau\ SM}}\right)^2 \\
 \frac{BR_{h\rightarrow WW}}{BR_{h\rightarrow WW\ SM}}=\left(\frac{g_{hWW}}{g_{hWW\ SM}}\right)^2
\end{IEEEeqnarray}
This is also the case of the vector boson fusion, VBF, process,
associated vector boson production and heavy quarks associated
production.
\begin{equation}
 \frac{\sigma_{ZH}}{\sigma_{ZH\ SM}}=\left(\frac{g_{hZZ}}{g_{hZZ\ SM}}\right)^2
\end{equation}
In the case of nested expansion or MSSM factorisable one, we have
used a modified version of \texttt{HDECAY} \cite{hdecay}. This
will be used for decays where susy loop contributions are not
negligible and also loop induced decays. It is for instance used
to compute $\Gamma_{h \to\gamma\gamma}$. Finally, observables
where no explicit loop computation was needed have been computed
with \texttt{CalcHEP} (\cite{calchep}) : this is the case for
instance of Higgs decay to other Higgses, where the loop
correction can be reproduced by an effective potential (see
\cite{boudjema_efflambda}).\\ Following this choice, several
approximations have been made. The most important is the gluon
fusion cross-section. As it is a nested expansion case, we started
with a modified version of \texttt{Higlu} (\cite{higlu}), an
MSSM-dedicated code. However, the integration over the parton
density functions set was an unacceptable lost in time,
considering that the effective parameter space was already
22-dimensional. Hence we used the approximation :
\begin{equation}
\sigma_{gg\rightarrow h}=\frac{\Gamma_{h\rightarrow gg}}{\Gamma_{h\rightarrow gg\ SM}}\sigma_{gg\rightarrow h\ SM}.
\label{ggh}
\end{equation}
Similarly if we wanted to consider SUSY loop corrections to the
Vector Boson Fusion (VBF), associated vector boson production and
heavy quark associated production, we should have used results
from MSSM-dedicated code, with the same issue on pdf. \\
About the branching ratios, using \texttt{HDECAY} did not
reproduce the best results for decays into off-shell gauge bosons,
but using a more precise tool such as \texttt{Prophecy4f}
(\cite{prophecy4f}) would have raised a lot the computation time,
since it goes up to event generation. At the end of the day,
production cross-sections were rescaled from the SM predictions
(avoiding hence the use of specific codes as \texttt{Higlu}, or
codes for VBF), and decays were rescaled from MSSM predictions
obtained with \texttt{HDecay} (avoiding the use of
\texttt{Prophecy4f}), which allowed for a large gain in computing
speed for a very moderate precision loss.

\subsection{Accuracy of the effective expansion}
 The parameters weighing the
 effective operators must be at most of order
 1, if we believe the high energy theory to be
 weakly coupled. Keeping the leading orders, in our case $1/M$ and $1/M^2$,
 should be sufficient and within the spirit of the effective approach. But even with this requirement we
 can still be confronted with situations in parameter space
where the expansion in $1/M$ can fail. If that is the case then
one expects neglected higher order terms such as $\grandO{1/M^n}$ with
$n>2$ to be non negligible. If the inclusion of these higher
order terms makes a difference, the expansion is not to be trusted
and such configurations of parameters should be discarded. In some
situations, for certain observables, the effect of the order $1/M,
1/M^2$ can be over amplified because the leading order
contribution is very small or that there is an accidental
suppression that makes the higher order effect important. For
 instance  when $M_{A_0}$ is low and $\tb$ moderate
 to high, then $h,H$ can get nearly degenerated at
 zeroth order in the effective expansion. Hence
 any observable involving
 $\sqrt{m_{H}^2-m_{h}^2}$ will have
 an ill-defined perturbative expansion, since the
 derivatives of the square root near 0 are
 infinite.\\

 Our check on the accuracy of
 the effective expansion is made on the light Higgs mass
 $m_h$, after all the reason for  including $\grandO{1/M^2}$
was because there were non negligible contributions from this
order to $m_h$. The masses of the CP-even Higgses are computed
from the Higgs
 $2\times 2$ mass matrix $\mathcal{M}$, which in
 the case where the Lagrangian is truncated at
 second order, writes as
\begin{equation}
\label{mhatM2}
 \mathcal{M}_{|2}=\mathcal{M}^{(0)}+\frac{c_5}{M}\mathcal{M}^{(5,1)}+\frac{c_6}{M^2}\mathcal{M}^{(6,1)}+\frac{c_5^2}{M^2}\mathcal{M}^{(5,2)}
\end{equation}
Note that $\mathcal{M}^{(0)}$ is the loop corrected mass matrix.
To have this concise form, we have used generic names for
effective coefficients : $c_5$ for the order 5 coefficients
$\{\zeta_{10},\zeta_{11}\}$ and $c_6$ for the $\{a_{ij}\}$. Now,
since $m_h$ is obtained by solving a quadratic equation, using
eq.~\ref{mhatM2} leads to a {\em solution}
that includes contributions up to $\grandO{1/M^4}$
\begin{equation}
 m_h=m_h^{(0)}+\frac{c_5}{M}m_h^{(5,1)}+\frac{c_6}{M^2}m_h^{(6,1)}+\frac{c_5^2}{M^2}m_h^{(5,2)}+\frac{c_5^3}{M^3}m_h^{(5,3)}+\frac{c_5c_6}{M^3}m_h^{\binom{5,1}{6,1}}+\grandO{1/M^4}
 \label{mh_an}
\end{equation}
So our first check on the effective expansion was to ensure that
the $\grandO{1/M^3}$ terms in eq. \ref{mh_an} were small compared to
the $\grandO{1/M^2}$ terms. We have therefore imposed the condition
\begin{equation}
 \left|\frac{\frac{c_5^3}{M^3}m_h^{(5,3)}}{m_h^{(0)}+\frac{c_5}{M}m_h^{(5,1)}+\frac{c_6}{M^2}m_h^{(6,1)}+\frac{c_5^2}{M^2}m_h^{(5,2)}}\right|+\left|\frac{\frac{c_5c_6}{M^3}m_h^{\binom{5,1}{6,1}}}{m_h^{(0)}+\frac{c_5}{M}m_h^{(5,1)}+\frac{c_6}{M^2}m_h^{(6,1)}+\frac{c_5^2}{M^2}m_h^{(5,2)}}\right|<0.1
\end{equation}
 so that points that do not pass this condition were discarded.\\

Once this algebraic test was passed, we performed another purely
numerical test based on the explicit inclusion of an operator of
$\grandO{1/M^3}$. Now the CP-even mass matrix becomes\footnote{The
reason why $c_5^3$ and $c_6c_5$ terms pop up is that the
Lagrangian is itself a non-linear function of the K\"ahler
potential and the superpotential.}
 \begin{equation}
 \mathcal{M}_{|3}=\mathcal{M}_{|2}+\frac{c_5^3}{M^3}\mathcal{M}^{(5,3)}+\frac{c_6c_5}{M^3}\mathcal{M}^{\binom{5,1}{6,1}}+\frac{c_7}{M^3}\mathcal{M}^{(7,1)},
\end{equation}
where $c_7$ stands for a new operator
 \begin{equation}
\mathcal{O}_7=\zeta_3\left(H_1\cdot H_2\right)^3.
\end{equation}
To compute the shift in $m_h$, we have run again \texttt{lanHEP}
including now the new operator, and requiring the Feynman rules to
be computed at order $1/M^3$. This being done we could evaluate
numerically $\mathcal{M}_{|3}$ and compute the resulting value for
$m_h$. To do this, we had to assign a value to the $c_7$
coefficient. We choose it to be the maximum (in absolute value) of
all lower-order coefficients.
$$c_7=max\left(|\zeta_{1l}|,|a_{ij}|\right).$$ The additional
constraint was set as
\begin{equation}
\left|\frac{m_h(\mathcal{M}_{|3})-m_h(\mathcal{M}_{|2})}{m_h(\mathcal{M}_{|2})}\right|<0.1
\end{equation}
Once again, a point failing these two constraints will be
discarded. Those two checks are complementary in the sense that
the first one ensures only that we do not hit any singular point
when computing the Higgs mass, which is essential to use
perturbation theory but does not say much about  the contribution
of higher-orders, while the second constraint is an explicit check
that the next order contribution is indeed small enough.

\section{Analysis}
We have fixed the supersymmetric parameters (except $\ma$ and
$\tb$), since we wanted to focus on the higher-order effects
rather than MSSM phenomenology. For the sake of concreteness, we
used the $m_{h max}$ scenario as defined in
\cite{carenaheinemeyer}. All soft masses are set to $M_{{\rm
soft}}=1$TeV, $\mu$ and $M_2$ are set to 300 GeV, $M_1$ is fixed
by the universal gaugino mass relation $M_1=\frac{5}{3}\tan^2
\theta_W M_2 \simeq M_2/2$, and $M_3=800$ GeV ($\cos^2
\theta_W=M_W^2/M_Z^2$). All trilinear couplings are set to 0,
except for $A_b=A_t=2M_{{\rm soft}}\frac{\mu}{\tb}$ that are set
to maximise the radiative corrections to $m_h$. Maximising $m_h$
through purely the MSSM part is not necessary since the new
effective operators will enhance the mass more efficiently. We
have also set $m_s=300$ GeV so that we have
\begin{equation}
 \frac{\mu}{M}=\frac{m_s}{M}=0.2
 \label{eq:ratio}
 \end{equation}
Those are, together with $v/M$, the
suppression factors of the effective
terms. The only free parameters have
been varied in the range
$$\tb\in[2,40]\qquad\ma\in[50,450].$$

Since we are dealing with a large parameter space (22
dimensional), performing a satisfactory scan is a crucial issue.
We attempted first a search with random Markov chains, but it
ended up to be limited by the frequentist character of the
technique : indeed a Markov chain will stay in regions depending
on the number of allowed points in it. Because our model exhibits
regions that are extremely more populated than others, the Markov
chains showed a tendency to stagnate. Besides, we are not
interested in the density of points, but in disentangling what
lies in the reach of such a generic susy model, and what is
incompatible with it. In particular one of our first motivations,
in accordance with the main reason for considering such scenarios,
is to explore regions where $m_h$ is much heavier than what is in
the usual MSSM. So after an exploratory random scan we aim at
populating regions giving largest values for $m_h$. We first carry
a blind random scan on all parameters, then we pick  up a point
exhibiting a large mass and scan again by perturbing around this
point. In fact we perturb only around the $\grandO{1/M^2}$
$a_{ij}$ values rescaling them by a common factor while rescanning
on the $\grandO{1/M}$ which give the leading order effect in the
increase of $m_h$. In the scans that we will show later, for
$\tb=2$, we perturb around
$$\begin{array}{|cc|cc|cc|cc|}
& & a_{10} & 0.168605 & a_{11} & -0.55814 & &\\
a_{12}& 0.511628 & a_{20} & 0.0465116 & a_{21} & 0.639535 & a_{22} & 0.802326\\
a_{30} & 0.151163 & a_{31} & 0.744186 & a_{32} & 0.284884 &  a_{40} & 0.238372\\
a_{41} & 0.383721 & a_{42} & 1. & a_{50} & 0.848837 & a_{51} & -0.133721\\
a_{52} & -0.732558 & a_{60} & 0.598837 & a_{61} & 0.575581 & a_{62} & 0.331395
\end{array}$$
We will refer to this combination as $c_1$. We have then carried
out reduced scans on the parameter space by randomly choosing a
triplet $(x,\zeta_{10},\zeta_{11})$ in the cube $[-1,1]^3$,
and associate it to the point
\begin{equation}
 p=(\zeta_{10},\zeta_{11},x\times c_1)
\end{equation}
This choice strongly relies on the fact that most of the
phenomenological change are brought by the order 5 operators, even
though the order 6 ones are essential in raising $m_h$ further.

\subsection{Characteristic features}
\begin{figure}[h!]
\begin{center}
\begin{tabular}{cc}
\includegraphics[scale=0.3,trim=0 0 0 0,clip=true]{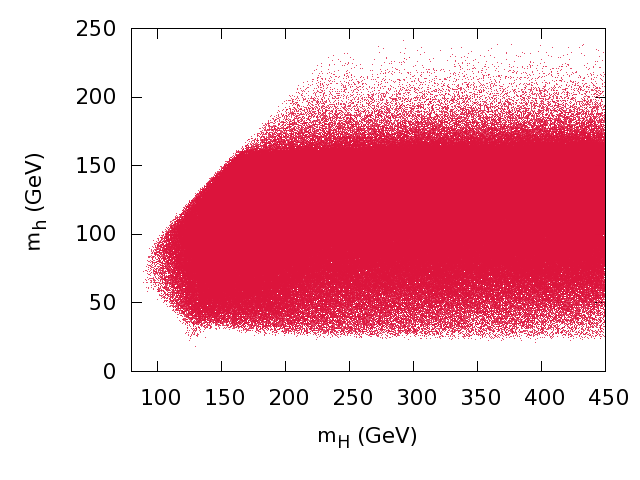}&
\includegraphics[scale=0.3,trim=0 0 0 0,clip=true]{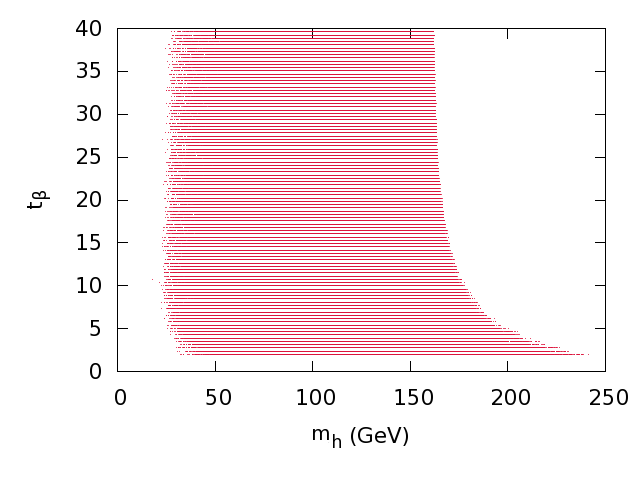}
\end{tabular}
\end{center}
\begin{center}
\begin{tabular}{ccc}
\includegraphics[scale=0.2,trim=0 0 0 0,clip=true]{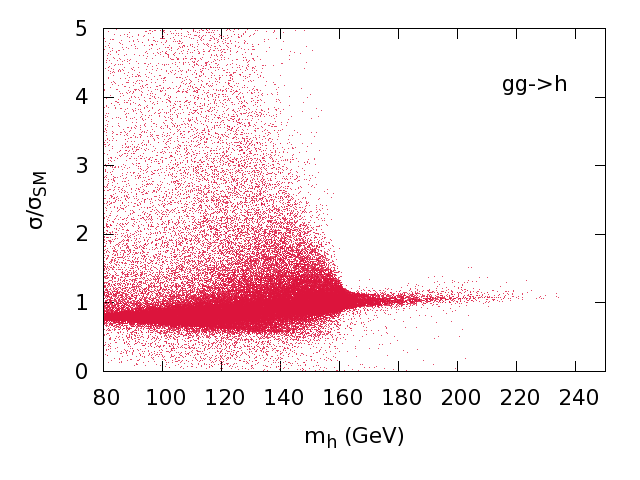}&
\includegraphics[scale=0.2,trim=0 0 0 0,clip=true]{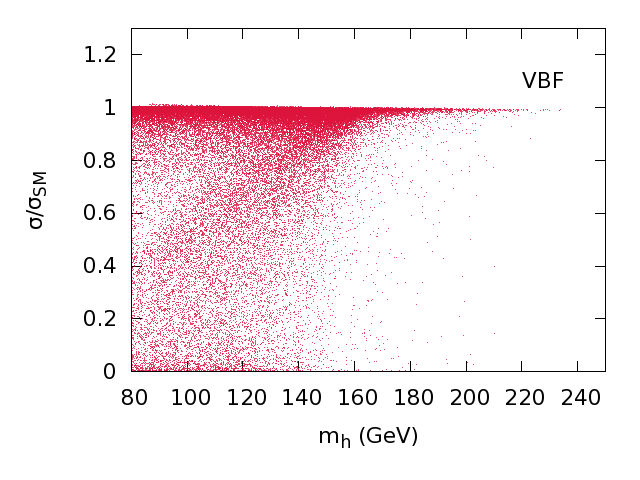}&
\includegraphics[scale=0.2,trim=0 0 0 0,clip=true]{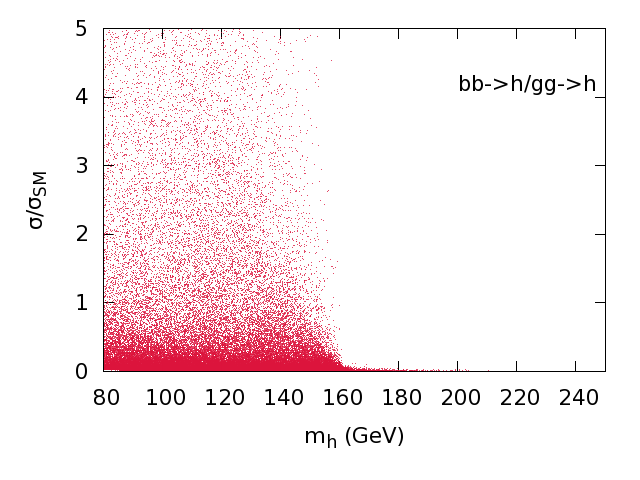}
\end{tabular}
\end{center}
\caption{\label{fig:pre_constraint} {\em Features of the BMSSM. In
all the plots no experimental constraint has been applied. In the
left-top pannel we show the range in the $m_H,m_h$ plane in the
BMSSM framework, then on the right the  $\tb,m_h$ dependence.
Plots in the second row all refer to the lightest CP-even Higgs.
We first show the cross section for $gg \to h$ normalised to that
of the SM as a function of $m_h$, then the similar normalised
cross section for VBF. The third plot shows the ratio
$\sigma_{bb\rightarrow h}/\sigma_{gg\rightarrow h\ SM}$ (this
ratio is more relevant than the normalised $bb\rightarrow h$
production since the $b$ quark fusion is rather low in the SM). }}
\end{figure}
Before applying the constraints coming from collider data, it is
worth briefly reviewing  what phenomenological changes the BMSSM
brings, compared to the SM or the MSSM case. As can be seen in figure
\ref{fig:pre_constraint}, the most characteristic feature is the
raise in the mass of the lightest CP-even Higgs up to 250 GeV. The
largest values can only be reached for low values of $\tb$, this
feature coming from the fact that the shift from dimension 6
operators is suppressed as $\tb$ increases. Note however that even
for $\tb=40$, the largest value of $m_h$ reaches about $160$ GeV, a
value larger by more than 20 GeV the corresponding value in the
MSSM. We note also some important changes in the production (and
the decays) of the light Higgs as compared to the SM case. We note
that gluon fusion can be drastically either enhanced or suppressed
especially for $m_h < 160$ GeV. As $m_h$ increases the changes in
this channel are less drastic, in particular the increase in $gg
\to h$ is very moderate. Of course, as we have just seen for $m_h
> 160$ GeV $\tb$ is low and therefore does not affect much $bbh$ which is enhanced
for high values of $\tb$. This is also clearly displayed in the
rate $bb \to h$ where the importance of this channel lies in the
region $m_h < 160$ GeV. In VBF one hardly finds an increase for any
Higgs mass, although for $m_h < 160$ GeV one sees many
configurations where the rate can be very small, one still finds
scenarios that point to the fact that even for $m_h=200$ GeV, the
lightest Higgs can be far from being SM-like.

\subsection{Indirect constraints from precision electroweak data}
We have seen in eq.~\ref{mwmz} that some operators lead to a
breaking of the custodial $SU(2)$ global symmetry. We have
implemented the constraints from precision electroweak observables following \cite{altarelli_epsilon}
using data from Ref.~\cite{lep_epsilon} and requiring agreement
within $2\sigma$. This leads to a very strong constraint on the
combination
\begin{equation}
 a_{10}-a_{30}\tb^2+a_{20}\tb^4
\end{equation}
as can be inferred directly from eq.~\ref{mwmz}. \\
We do not consider constraint from the flavour sector or even baryogenesis in the BMSSM \cite{blum_bmssmbaryogenesis_0805,blum_bmssmbaryogenesis_1003,bernal_bmssmflavour_1104,altmannshofer_bmssmcp_1107} as one has to specify the details
of the matter sector. One can easily  bypass possible constraints from this sector by changing the details of the model.
\subsection{Constraint from $t \to H^+  b$}

Constraints from not too heavy charged Higgses are imposed
exploiting the search of charged Higgs boson in top decays, done
by CMS (\cite{cms_eps_Hp}). The latter explores the channel
$t\rightarrow H^+ b$, $H^+\rightarrow\nu\tau^+$. Special care was
taken in the computation of the branching ratio of $t\rightarrow
H^+ b$, since it can be affected both by QCD corrections and by
supersymmetric-QCD corrections. The first have been included using
the \texttt{HDecay} code, and the second by including the $\Delta
m_b$ correction following\cite{deltamb_spira,micromegas2}. To end
up with the correction branching ratio, QCD corrections were also
taken into accounts for $t\rightarrow W^+ b$ using
\texttt{HDecay}.

\subsection{Higgs collider searches: generalities}
The Higgs direct searches at the colliders are taken into account
by comparing the ratio $\sigma/\sigma_{SM}$ at the 95\% CL
exclusion value for each analysis by LEP, TEVATRON and the LHC.
This is automated via HiggsBounds \cite{HB} for LEP, Tevatron and
some of the LHC results.\newline For the neutral Higgs bosons, we
must account for the case where $h$ and $H$ get degenerate : in
this case the two cross-sections must be added. We define two
Higgs bosons to be degenerated when their mass difference is less
than 10 GeV for hadron colliders (LHC and Tevatron) and 2 GeV for
LEP. The implementation of LHC results from Lepton Photon 2011 had
to be done separately, as we show now.

\subsection{LHC implementation of the neutral Higgs searches}
\subsubsection{An ``inclusive" analysis}
\label{subsub-incl}
\begin{figure}[h!]
\begin{center}
\includegraphics[scale=0.3,trim=0 0 0 0,clip=true]{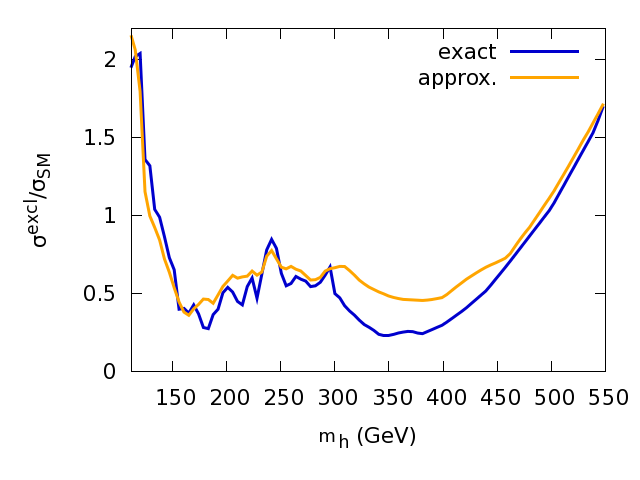}
\end{center}
\caption{\label{fig:sm_comb} {\em The figure shows the excluded
cross sections for the SM Higgs production at the 95\% confidence
level in the case of the combined analysis performed by the ATLAS
and CMS collaborations, exact, compared to the approximate
combination we perform based on the quadrature sum, approx.}}
\end{figure}

In this first analysis all Higgs search data from from both ATLAS
and CMS as presented at {\em Lepton Photon 2011} are used ranging from
$1$fb$^{-1}$ to 2.3 fb$^{-1}$, that we will sometimes refer to for short as $2$fb$^{-1}$ data,  that is to say :
\begin{itemize}
 \item $H\rightarrow\gamma\gamma$, done by ATLAS (\cite{atlas_lp_gamgam}) and CMS (\cite{cms_lp_gamgam}).
 \item $VH\rightarrow V\oo{b}b$, done by CMS (\cite{cms_lp_bb}).
 \item $H\rightarrow WW$, done by ATLAS on different final states ($l\nu l\nu$ \cite{atlas_lp_ww}, $l\nu qq$ \cite{atlas_lp_wwb}) and CMS ($l\nu l\nu$ \cite{cms_lp_ww}).
 \item $H\rightarrow ZZ$, done by ATLAS on different final states ($4l$ \cite{atlas_lp_zz}, $2l2q$ \cite{atlas_lp_zz_2l2q} and $2l2\nu$ \cite{atlas_lp_zz_2l2nu}) and CMS ($4l$ \cite{cms_lp_zz}, $2l2q$ \cite{cms_lp_zz_2l2q}, $2l2\nu$ \cite{cms_lp_zz_2l2nu} and $2l2\tau$ \cite{cms_lp_zz_2l2tau}).
 \item $H\rightarrow\tau\tau$, done by ATLAS (\cite{atlas_lp_tautau}) and CMS (\cite{cms_lp_tautau}).
\end{itemize}
A priori, all these analyses are dedicated to the SM, however we
can still try to compare the cross-section we obtain in the BMSSM
model with the excluded cross-sections from the collaborations. To
this aim we will add all different production cross-sections :
gluon fusion ($gg\rightarrow h$), $b$ quark fusion ($bb\rightarrow
h$), vector boson fusion (VBF) and associated vector boson
production (VH). All SM cross-sections have been taken from the
LHC Higgs cross-section working group
(\cite{LHC_Higgs_cs1,LHC_Higgs_cs2}) except for the $b$ quark
fusion, computed with \texttt{bbh@NNLO} (\cite{bbh_at_nnlo}).\\
After computing the ratios $\sigma/\sigma^{\text{excl}}$ for each
channel, we combine all channels by adding the ratios in
quadrature. Moreover, for analyses that exist only in one of the
collaborations ($V b \bar b$ in CMS, $WW\to l\nu qq$ ATLAS, $ZZ
\to ll \tau \tau$ in CMS), we make up for the lack of the
corresponding analysis by including it in our analysis through a
scaling factor $\sqrt{2}$ to the corresponding ratio. This
approach is followed in Ref.\cite{carena_bmssmhiggs_1111} also.
The test applies separately to   the three neutral Higgses (though
in the CP-odd case, some analyses like $H \rightarrow WW$ do not
apply) and rejects all points were at least one Higgs fails to
pass the test.  The excluded {\em inclusive} cross section
obtained by such an approximation on a SM boson is shown in fig~
\ref{fig:sm_comb}. \\
Since the exact combination of ATLAS and CMS channels in the SM
case has recently been published \cite{atlas_cms_lp_comb}, we have
presented it on the same plot, fig~ \ref{fig:sm_comb}, so one can
quantitatively weigh the discrepancy coming from the quadrature
sum approximation.
Fig.~\ref{fig:sm_comb} shows, that apart from the range
$300<m_h<450$ GeV where all $ZZ$ channels combine, the
approximation based on the quadrature sum is well justified. In
our model, this range is not reached by the lightest Higgs nor by
the CP-odd Higgs which does not couple to vector bosons. Moreover
even in the BSSM for $m_H>300$ GeV, one is most often in the
decoupling limit where the $HWW/HZZ$ is vanishingly small.
Therefore for the BSSM the approximate combination should be
trustworthy.\\

\begin{figure}[!h]
\begin{center}
\begin{tabular}{cc}
\includegraphics[scale=0.3,trim=0 0 0 0,clip=true]{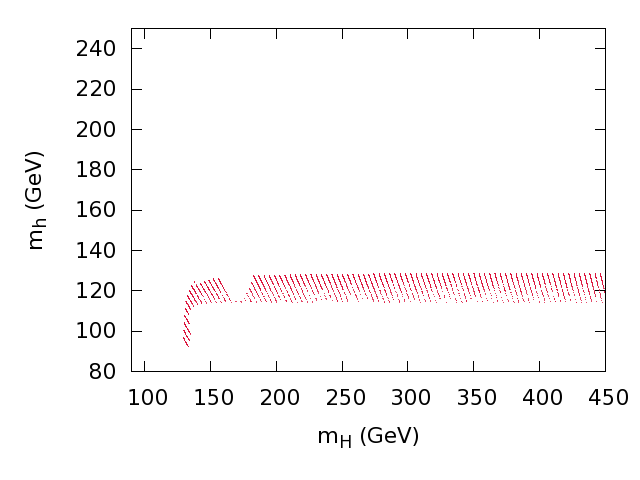}&
\includegraphics[scale=0.3,trim=0 0 0 0,clip=true]{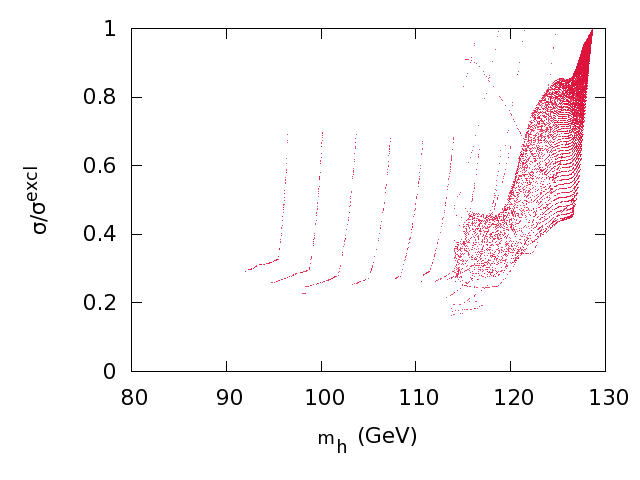}
\end{tabular}
\end{center}
\caption{\label{fig:mh_MSSM} {\em The allowed range in the
$m_h-m_H$ plane for our reference MSSM model is shown in the left
panel. The right panel shows the ratio $\sigma/\sigma^{{\rm
excl.}}$ as a function of $m_h$.  }}
\end{figure}

\begin{figure}[h!]
\begin{center}
\begin{tabular}{cc}
\includegraphics[scale=0.3,trim=0 0 0 0,clip=true]{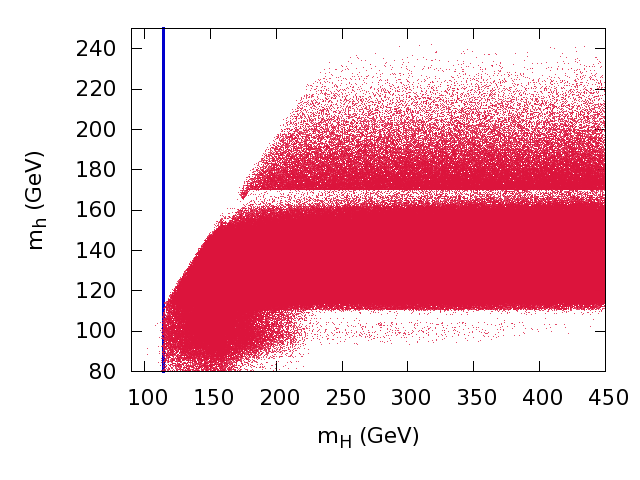}&
\includegraphics[scale=0.3,trim=0 0 0 0,clip=true]{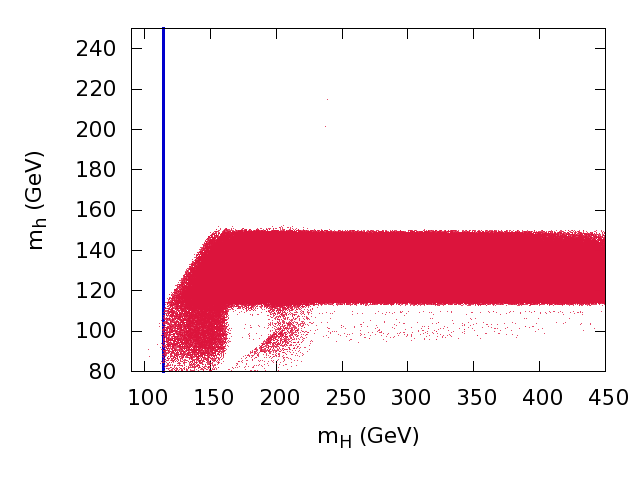}
\end{tabular}
\end{center}
\caption{\label{fig:mh_BMSSM} {\em BMSSM predictions before (left)
after (right) applying the LHC constraints in the $(m_H,m_h)$
plane. The vertical blue line shows the LEP SM bound $m_\Phi=114$
GeV.}}
\end{figure}

We show in fig.~\ref{fig:mh_MSSM} and \ref{fig:mh_BMSSM} the
allowed points obtained with the analysis we have just described,
either in the MSSM and BMSSM cases. In the MSSM case,
fig.~\ref{fig:mh_MSSM}, the light Higgs mass is distributed in
between the LEP bound (114 GeV) and the maximum of the radiative
corrections (about 130 GeV). We have also plotted here the ratio
$R_\sigma=\sigma/\sigma_{\text{excl}}$ of each point (it is not
necessarily a $h$ signal, but can be any of the three Higgs
bosons) against the mass of the lightest Higgs. We notice that
this MSSM scenario is largely unaffected by the current results
since the ratio between the predicted production rate to the
excluded production rate goes down to 0.2, so that one would
require an increase by a factor 25 in the luminosity to exclude
this particular MSSM model. Models with the highest $m_h$
predicted in this model (up to 130 GeV) require much less
luminosity increase to be excluded or discovered.
\\

In the case of the BMSSM, the fact that $m_h$ can be raised to as
high as $250$ GeV changes the picture quite drastically compared to
what was allowed before the LHC data (LEP and Tevatron data are
included in both sets). Fig.~\ref{fig:mh_BMSSM} shows that with
just about $2$fb$^{-1}$ of collected data the $m_h-m_H$
plane has shrank considerably due to the fact that a rate 2 times
smaller than the SM for $m_h>160$ GeV is excluded. This shows in
particular that $m_h > 150$ GeV is now excluded. Therefore the main
{\em raison d'\^{e}tre} of such models that aimed at raising the
lightest Higgs mass considerably is now gone. Only an extra
$15$ GeV increase for the lightest Higgs compared to the maximal
value attained in the usual MSSM framework is still allowed. Therefore the majority
of models that survive have $114<m_h<150$ GeV, but we do find some regions with smaller values of $m_h$.\\

Indeed, while we find that the heaviest CP-even Higgs is above the LEP
limit, $m_H>114$ GeV, in the range $114 < m_H < 220$ GeV we find
models where the lightest Higgs is lighter than the LEP limit of $114$ GeV, we even find that models with $m_h<M_Z$
are still possible. In these configurations the lightest Higgs is far
from being SM-like. We have seen in fig.~\ref{fig:pre_constraint}
that the $hWW$ coupling can be drastically reduced. In this
case it is $H$ that picks up almost the totality of the $HWW/HZZ$
coupling, which explains why $m_H>114$ GeV (LEP constraint). The
configuration with $m_h<100$ GeV consists of two separate scenarios as
fig.~\ref{fig:mh_BMSSM} shows. One notices a region that
corresponds to $m_H> 2 m_h$ starting at $m_H=160$ GeV. Here the
branching ratio $H\mapsto hh$ can be as high as 0.6, with $h$
decaying almost exclusively to $b$ quarks, makes such scenarios
difficult to probe at the LHC. For $114< m_H < 160$ (GeV), some
scenarios are still viable because they correspond to $gg \to H$
that can go down to $50\%$ the value of the SM. Since this
reduction is limited to no more than $50\%$, such scenarios will
eventually be excluded by a luminosity increase. Other scenarios
in this mass range have a $\tb$ enhanced $bbH$ coupling, which
is constrained through $VH\rightarrow V\oo{b}b$ and
$H\rightarrow\tau\tau$ which includes $b\bar b \to H$ (since the BR to $\tau$ is also
significantly enhanced). As the luminosity will increase so will
the sensitivity of these last two channels.

Ref.~\cite{carena_bmssmhiggs_1111} has very recently carried a
very similar analysis. Our results are in very good agreement with
theirs apart from the region with $m_h<100$ GeV which is more
populated in our case. Note that although we carry a similar LHC
analysis, we differ in the choice of the MSSM reference point and
more importantly in the scan over parameters. One example is the
scan in $\tb$ that is covered uniformly in the range $2$ to $40$
in our case, whereas in Ref.~\cite{carena_bmssmhiggs_1111} the
emphasis was on $\tb=2$ and $\tb=20$  with a sparse scan in
between. In fact we have verified that our models that survive the actual LHC Higgs constraints have
$\tb$ in the range $5$ to $15$.

\subsubsection{Refining the analysis: lessons from a fermiophobic Higgs}
The analysis we have just performed, so called inclusive, has
exploited results tailored for SM Higgs searches to constrain a
model with properties that are quite different from those of the
SM as fig.~\ref{fig:pre_constraint} makes evident. Therefore the
bias such an approach introduces can be questioned considering
that the weight of the different production mechanisms can be very
different from those of the SM. An exclusion given on the
inclusive $\sigma\times BR$ quantity implicitly assumes that the
ratio between each production mode is the same as in the Standard
Model, in which case the ratio of exclusive cross-sections is
identical to the ratio of inclusive cross-sections. However this
assumption breaks down on many BMSSM points. In a nutshell, the
applicability of the SM search depends on whether the analysis can
differentiate between the different production modes of the Higgs,
which would allow to fold in  the weight of the different channels
in the analysis. Exclusive searches are performed by the
collaborations with the aim of selecting a particular sub-channel
to enhance sensitivity to a model that would be diluted in a more
inclusive search. Analyses  for $H \to WW$ that define three
categories $H$+0 jet, $H$+1 jet and $H$+2 jet aim for example at
enhancing, with the 2-jet,  the VBF contribution. The 0-jet
subchannel will be quite sensitive to the gluon fusion, since the
main part of its inclusive cross-section falls into this category,
but very slightly to the VBF. Therefore if one takes a model with
gluon fusion dominating all other production modes, like a heavy
4th generation (and to a certain extent the SM), the exclusion
will be purely driven by the 0-jet subchannel. If one takes a
fermiophobic model, the gluon fusion vanishes  and the exclusion
is given by the 2-jets subchannel. It is clear that the signal
obtained in the 4th generation model or the fermiophobic model
with the same inclusive cross-section (combining 0,1,2-jet) is not the same and will not
lead to the same exclusion limits. In such a case one would then
have to scale the cross-section reweighted by the efficiency of
each mode ($gg$, VBF, ...) that contributes to the same inclusive
final state, that is to say computing the ratios that quantify the exclusion power

\begin{equation}
 R_{\text{incl}}=\frac{\sigma}{\sigma_{SM}}=\frac{\sum_i\sigma_i}{\sum_i\sigma_{SM\
 i}} \quad\longrightarrow\quad
 R_{\text{excl}}=\frac{\sum_i\sigma_i\epsilon_i}{\sum_i\sigma_{SM\
 i}\epsilon_i} \label{eq:comp}
\end{equation}
where $i$ runs over the production mode (gluon fusion, VBF and so
on) and $\epsilon_i$ is the associated efficiency. We will then
compare with the excluded value. One can estimate the difference
between the two approaches of eq. \ref{eq:comp} by plotting the
excluded cross-section of different models for a single search.
\\

\begin{figure}[h!]
\begin{center}
\begin{tabular}{cc}
\includegraphics[scale=0.3,trim=0 0 0 0,clip=true]{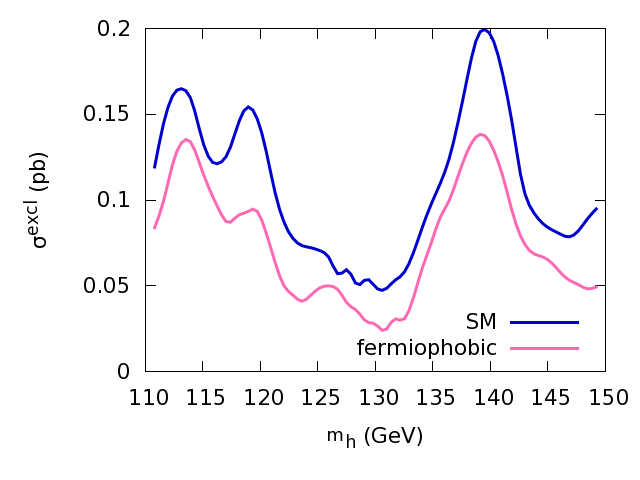}&
\includegraphics[scale=0.3,trim=0 0 0 0,clip=true]{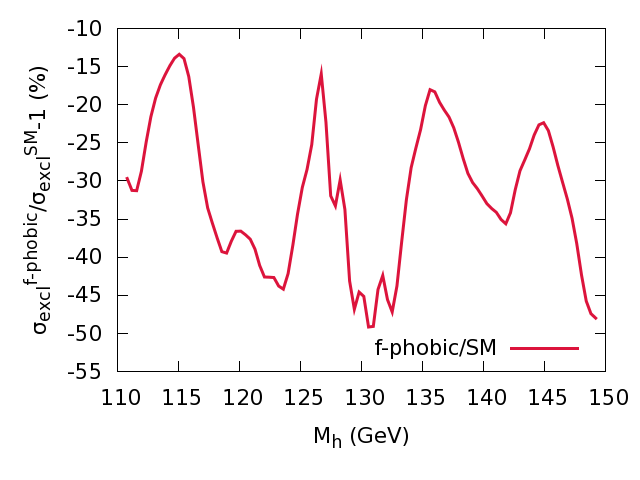}
\end{tabular}
\end{center}
\caption{\label{fig:gamgam} {\em The panel on the left shows the
SM and fermiophobic excluded cross-sections in the $H\rightarrow
\gamma\gamma$ CMS analysis, these plots are extracted from
Ref.~\ref{fig:gamgam}. Cross sections are given  in picobarns. On
the right is shown the relative difference between fermiophobic
and SM analyses, in percent units.}}
\end{figure}

For $H \to \gamma \gamma$ a category with a harder $p_T^{\gamma
\gamma}$ favours Higgs production through VBF or vector boson
associated production over $gg$ induced where the Higgs has a
smaller $p_T$. Such separation could very useful and efficient in, again,
the case of a fermiophobic Higgs whose $gg$ induced cross section
is vanishing, in sharp contrast to the SM Higgs. This particular
model can be used as an example, though perhaps extreme since
one important SM channel is absent, to quantify the difference one
gets from an inclusive (in this case merging all $p_T^{\gamma
\gamma}$ regions)   compared to an exclusive search or
exclusion limit  for each $p_T^{\gamma
\gamma}$ region. Such an approach has been performed by CMS
\cite{cms_lp_gamgam}\footnote{Note that very recently a similar
analysis was also released by the ATLAS collaboration
\cite{atlas_hcp_gamgam}, which we however do not consider here}.
In that analysis the classification is done according to
$p_T^{\gamma \gamma}>40$ GeV for enhancing the fermiophobic signal
over background. If we consider the inclusive cross-section, in
this case no $p_T^{\gamma \gamma}$ separator, to set the limits on
$\sigma\times BR$, the limit would be the same in any model since
it would only be computed from the measured number of events, the
predicted background and its systematics (if we neglect
theoretical systematics). Using the exclusive method based on the
$p_T^{\gamma \gamma}>40$ GeV as a separator gives a much more
powerful limit, though model dependent limit as shown in fig.~\ref{fig:gamgam}.
Unfortunately CMS does not provide separate exclusion limit for each region in $p_T^{\gamma \gamma}$.
In fact CMS gives the values obtained for $\sigma^\text{excl}$ using both models, in the merged (inclusive) and the split (exclusive) analyses for  $m_h=120$ GeV.
$$\begin{array}{ccc}
 & \sigma_\text{merged}^\text{excl} (pb) & \sigma_\text{split}^\text{excl} (pb)\\
 \text{SM} & 0.1308 & 0.1104\\
  \text{fermiophobic} & 0.1303 & 0.0696\\
\end{array}
$$
One notices first that the values of $\sigma_\text{merged}^\text{excl}$ are nearly the same in the two models, which emphasize the fact that inclusive limits are more model independent than exclusive ones.  We note that the gain in the SM is approximately 20\%, and in the fermiophobic model nearly 50\%. Therefore applying
this $20\%$ correction to the inclusive analysis mimics the effect of a more refined two-region analysis.

This mean that if one had used the SM limit with the inclusive
approach in the context of a fermiophobic Higgs, one would have
lost a factor 2 in sensitivity compared to a more refined
optimised exclusive analysis.\\
Unfortunately at present the details of the analyses performed by
ATLAS and CMS do not provide all the needed information and
efficiencies that we require for an exclusive approach. At present
in a phenomenological analysis like ours the best that can be done
is to simulate the experimental analysis through a Monte-Carlo with the caveat
that some detector issues contributing to the efficiencies may be lost.

\subsubsection{Refining the analysis}
Improving the analysis means that we will attempt to exploit those channels where separators leading to exclusive observables
have been conducted. Of course the situation is different from the case of the fermiophobic model in the $\gamma \gamma$ signature
where only one channel is selected. Moreover the fermiophobic model is well defined, $gg \to h$ cross section is vanishing. In the scans we perform in the BMSSM case one is in fact considering many models where  a given Higgs mass, $m_h$,
corresponds to models with very different properties. Let us first go through all the channels we have used in the previous analysis and comment on how
one could, for some of them, take into account the exclusive nature of a particular final state.

\begin{figure}[h!]
\begin{center}
\includegraphics[scale=0.3,trim=0 0 0 0,clip=true]{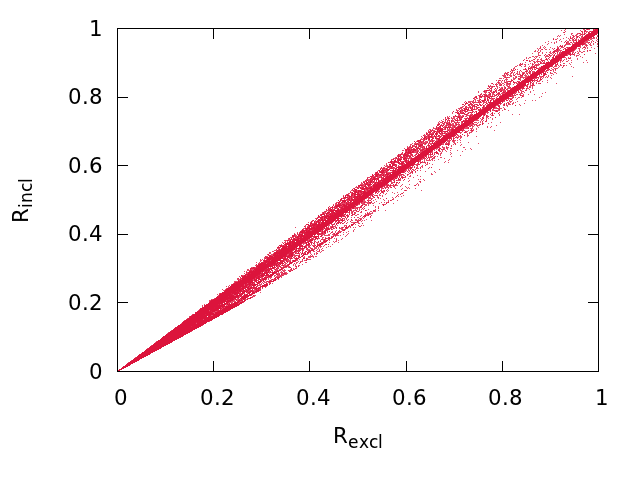}
\end{center}
\caption{\label{fig:ww_incl} {\em We show the exclusion power based on the inclusive analysis compared to the refined analysis, see text, applied to searches for the heaviest CP-even Higgs}}
\end{figure}

\begin{itemize}
 \item $VH\rightarrow V\oo{b}b$. In this case there is only one
 production mode, the vector boson associated production. Although
 it is strictly speaking two modes, the $Z$ and $W$, the scaling
 factor from the SM is nearly exactly the same, which simplifies
 the analysis. Here we can safely use inclusive cross-sections,
 since we only rescale the SM production.

\item $H\rightarrow \tau^+\tau^-$.
This channel is of interest in the MSSM and BMSSM for high $\tb$. $H$ is produced
either through $gg$ fusion of $bb$ fusion.
The ATLAS analysis (\cite{atlas_lp_tautau}) presents excluded cross-sections
for each of these two production modes.  This is most useful when analysing a new model as we can weigh each sub-channel separately.  This piece of information is extremely helpful since
it gives the efficiency in a very handy way : one has just to
compute the ratio of each production cross section to its excluded
value, sum them and compare to 1. Indeed as we deal with a
counting experiment, this is adding events from each production
mode and compare with the excluded number of events, which, in the
approximation of no theoretical systematics, is justified.

 \item $H\rightarrow ZZ\rightarrow 4l$.
Unlike the $WW$ signature where an analysis including 0-jet, 1-jet and 2-jet is performed, for the $ZZ$ channel one only has at the moment
a fully inclusive analysis.

\item $H\rightarrow \gamma\gamma$.
We have just seen in the fermiophobic Higgs search that CMS, and similarly ATLAS, divide the phase space
according to $p_T^{\gamma \gamma}$ that allows to give different exclusion limits if one assumes a fermiophobic
model compared to the SM.   As we have just argued,  the efficiencies in the two regions are not given. Our procedure here is to correct
the inclusive analysis of CMS \cite{cms_lp_bb} by $20\%$ to  recover the fully inclusive limit. Although this scaling was derived for $m_h=120$ GeV,
considering the narrow range of the $\gamma \gamma$ channel we assume this scale factor to be roughly constant. This is a conservative approach, but a precise analysis requires the exclusion cross section for each subchannel (here the $p_T^{\gamma \gamma}$ regions) and the efficiencies of each mode.

 \item $H\rightarrow WW\rightarrow l\nu l\nu$. Both ATLAS and CMS split the channel according to the number of
 recorded jets, which allow to gain sensitivity to specific
 production modes ($gg \to H$ or VBF). Fortunately enough, ATLAS provides exclusion
 limits for the 0-jet and 1-jet subchannel (and should soon provide  the 2-jet bin). Providing the 2-jet that would select the VBF would
 be extremely useful. Once again though the weight of the 0-jet and 1-jet in the ATLAS analysis are folded in, these weights are not provided.
Simulating the ATLAS analysis one could in principle calculate these weights or efficiencies. We have run
\texttt{PYTHIA} for a SM Higgs boson through
gluon fusion, VBF or $b$ quark fusion and extracted  the
efficiency of each production mode. Although this may seem far too
naive since full detector simulation is not applied we are only interested in the relative efficiencies, say
the ratio between the VBF  and gluon fusion. One  expects that a full detector simulation does not affect these ratios much.
The ratios we calculated were validated by the ATLAS collaboration\footnote{The VBF ratio to gluon fusion was in
very good agreement. Private communication.}. $b$ fusion could not be checked since it is not included in a SM Higgs analysis.  We were then able to fold in these ratios
witin a refined exclusive analysis.  We show in fig
\ref{fig:ww_incl} the relative difference between the
inclusive and exclusive, defined in eq~\ref{eq:comp}. This
relative correction is mainly positive, up to $30$\% which can be
traced back to the fact that the $b$ fusion efficiency is higher
than the gluon fusion one.
\end{itemize}

\begin{figure}[h!]
\begin{center}
\begin{tabular}{cc}
\includegraphics[scale=0.3,trim=0 0 0 0,clip=true]{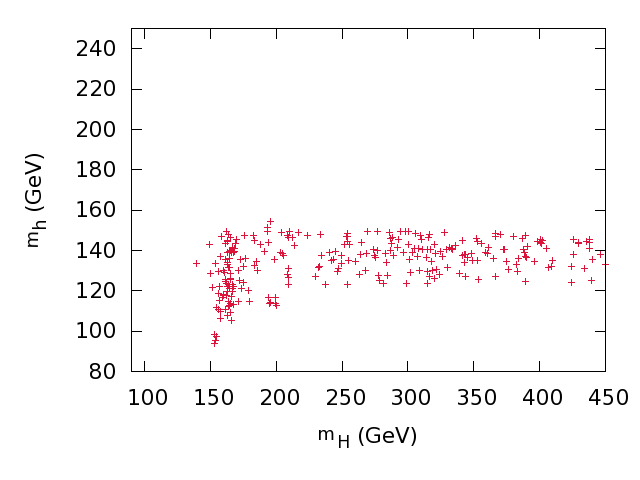}&
\includegraphics[scale=0.3,trim=0 0 0 0,clip=true]{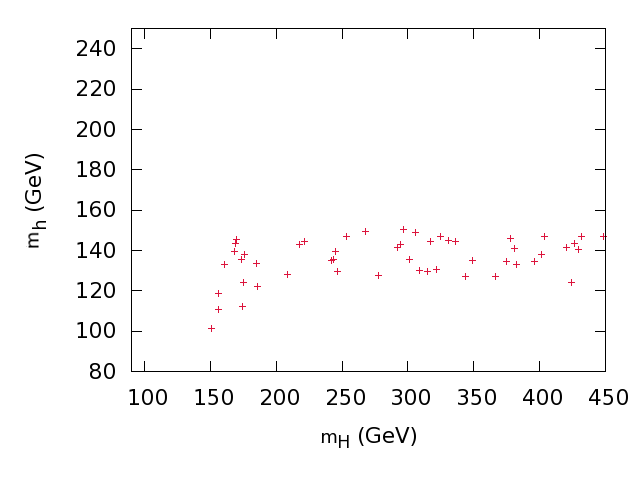}
\end{tabular}
\end{center}
\caption{\label{fig:mh_subset} {\em Taking a small subset of models, we apply an inclusive analysis, left panel, and compare it to the result of a refined analysis, right panel. }}
\end{figure}

To summarise we see that for the moment, the refinement concerns only two channels and may seem a modest improvement, but it is important to send a signal and a request  to the collaborations so that details of the analyses with the weight and efficiencies of all channels and sub-channels be released. It is important to stress that what we call the refined analysis is our approach to arrive at what we think is a better treatment of such models, nonetheless with the inclusive analysis this allows to compare and quantify the assumptions. It should also be clear that the refined analysis does not necessarily mean that it is more constraining than the inclusive one.
Before turning
to the final results taking into account these refinements and in order to understand their impact when scanning over a large set of parameters, we compare the exclusion power in terms of the inclusive approach compared to the refined analysis, eq.~\ref{eq:comp}, applied to the heaviest CP-even Higgs for illustration. The comparison is shown in fig.~\ref{fig:ww_incl} corresponding to the luminosity $2$fb$^{-1}$ (Lepton-Photon 2011). Note that we only display values with $R_{{\rm incl}}<1, R_{{\rm excl}}<1$ corresponding to models that are still viable. When the luminosity increases, the condition $R_{{\rm incl}}<1, R_{{\rm excl}}<1$ can be read from the plot, but would correspond to smaller $R$ values.
~
The figure shows that there is, unfortunately, little spread around $R_{{\rm incl}}=R_{{\rm excl}}$, the largest differences attaining about $20\%$ for $R<0.3$.  A scan over the entire parameter set, taking all constraints on all Higgses, showed practically not much difference between the refined and inclusive approach when projected on the $m_h-m_h$ plane. So we will not show such plots. However to illustrate that the two analyses do exclude different sets of models, we have generated a well chosen subset of models\footnote{The subset has $R_{{\rm excl}}>0.99$ applied  to all three Higgses. In the refined analysis $R_{{\rm excl}}<1$ is imposed while in the inclusive analysis $R_{{\rm incl}}<1$.}
 and passed them through the two  analyses, inclusive and refined. In this (biased) chosen subset of models, we see in fig.~\ref{fig:mh_subset} that the refined analysis excludes many more models.  Had we performed a full scan, the differences in the projection on the plane $m_h-m_H$ will hardly be visible.

\subsection{Expectations with higher luminosities}
\begin{figure}[h!]
\begin{center}
\includegraphics[scale=0.3,trim=0 0 0 0,clip=true]{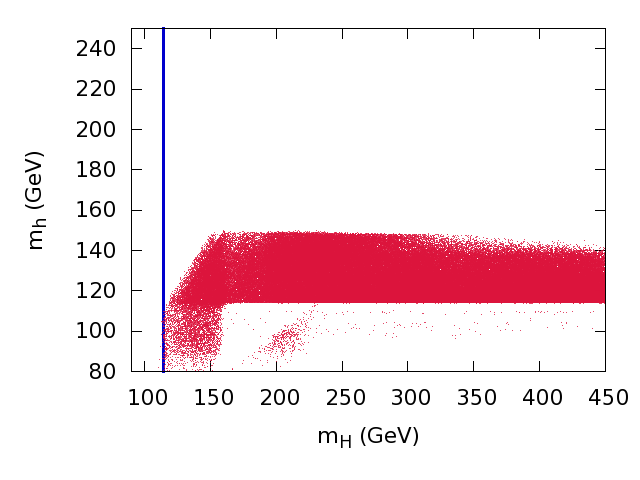}\\
\begin{tabular}{cc}
\includegraphics[scale=0.3,trim=0 0 0 0,clip=true]{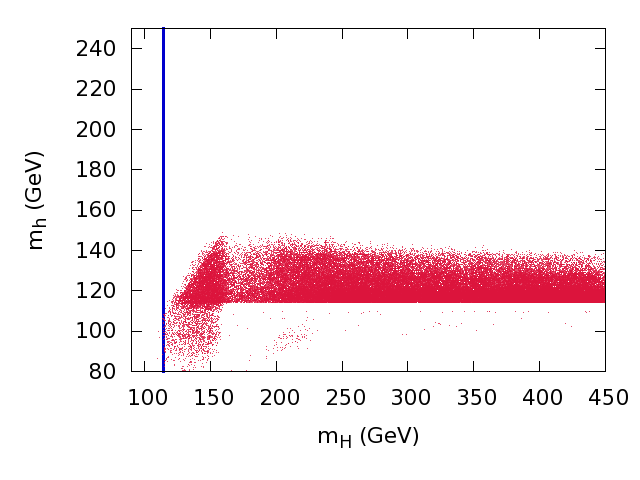}&
\includegraphics[scale=0.3,trim=0 0 0 0,clip=true]{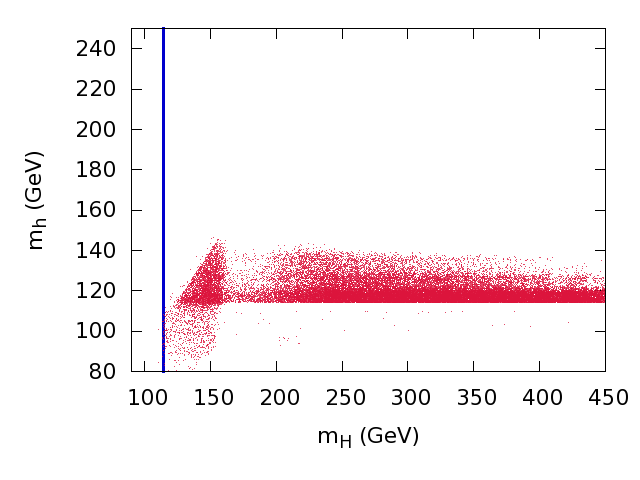}\\
\end{tabular}
\end{center}
\caption{\label{fig:expect} {\em Remaining parameter space when the luminosity increases. The top panel shows the expectation for 5fb$^{-1}$. In the second row the panel on the left is for 10fb$^{-1}$and the one on the right is for 15fb$^{-1}$.}}
\end{figure}

The LHC is running wonderfully. We have seen that the first set of data on Higgs searches released this summer has already put severe constraints on models such as the BMSSM. It is therefore very tempting to make projections with an increase of luminosity. Indeed ATLAS and CMS will very soon release data with $5$fb$^{-1}$. We show in fig.~\ref{fig:expect}, by simply rescaling our refined analysis by the luminosity factor, how the picture changes as the luminosity is increased from $5$, $10$ to $15$fb$^{-1}$. \\
Compared with the limits based on the released data with
$2$fb$^{-1}$, we notice first that as the luminosity increases the
upper limit on $m_h$, which sits now at $150$ GeV, decreases
gradually affecting first those models associated with $m_H>300$
GeV. Already with $10$fb$^{-1}$, there are practically no models
with $m_h>140$ GeV for $m_H>300$ GeV. With $15$fb$^{-1}$ only very
few models with $m_H < 150$ GeV remain. With this luminosity the
number of models with $m_h<114$ GeV is also drastically reduced
and at $15$fb$^{-1}$ they are practically all gone if $m_H>150$
GeV. At this luminosity, though much reduced, the  space with
$m_h<114$ GeV and $m_H<150$ GeV is still there. This said at this
luminosity, if the density of surviving models is any indicator
for the likelihood of a model, we see that there is a
concentration of  models within a thin layer with $114<m_h<120$
GeV when $m_H>250$ GeV, although of course many models still
survive with $m_h$ up to $m_h=140$GeV.

\subsection{Higgs decays to invisibles}
We present now the preliminary results of the consequence of the
Higgs decaying to invisible particles\footnote{The earliest
mention  of an invisible Higgs and its connection to dark matter
that we are of is made in a simple extension of the standard model
\cite{zee_darkons_1985}.}. Despite numerous advantages, the $m_{h\
max}$ scenario does not cover the full diversity of the MSSM nor
the BMSSM, in particular it does not cover cases where the Higgs
can decay to neutralinos, in particular the lightest ones. The
latter are good dark matter candidates and therefore these decays
of the ligtest Higgs are into invisibles. In order to have a
sufficient branching ratio to the neutralino one must have a
neutralino which is light enough, $M_{\tilde{\chi}^0_1}<m_h/2$. We
do not wish here to conduct a thorough analysis of the BSSM
higgses into invisibles and review all the constraints from dark
matter, we leave this to a more focused study. Dark matter issues
within the BSSM taking into account the dim-5 operators were
conducted in
\cite{cheung_bmssmdm_0903,gondolo_bmssmdm_0906,bernal_bmssmcosmo_0906,goudelis_bmssmdm_0912}.
Though succinct our implementation includes dim-6 operators
automatically. In the recent approach of
\cite{romagnoni_bmssmgoldstino_1111} which can be related to a
BSSM implementation, decays are into invisible light scalars.

In this exploratory study we consider $M_{\tilde{\chi}^0_1}<80$
GeV. In order to achieve this while taking into account LEP limits
on the chargino mass, such light neutralino are dominantly
bino-like. However in order to couple to the Higgs efficiently
there must be a higgsino component that is not too negligible, see
for example \cite{boudjema_higgs_inv}. One should therefore have
$M_1$, the bino mass, and $\mu$ not too far apart. We will set
$M_1=50$ GeV to have a light neutralino and $\mu=200$ GeV to have
enough mixing. The alert reader will have noticed that this value
of $\mu$ is smaller than what we have been using so far. In order
that our previous results are unaltered so that we can compare
with what an invisible decay brings, one should remember  that the
phenomenology without invisibles is unchanged if one keeps the
ratios $\mu/M,\ m_s/M$, that governs the effective expansion in
$1/M$, identical to what
was stated in eq.~\ref{eq:ratio}. \\

\begin{figure}[h!]
\begin{center}
\includegraphics[scale=0.4,trim=0 0 0 0,clip=true]{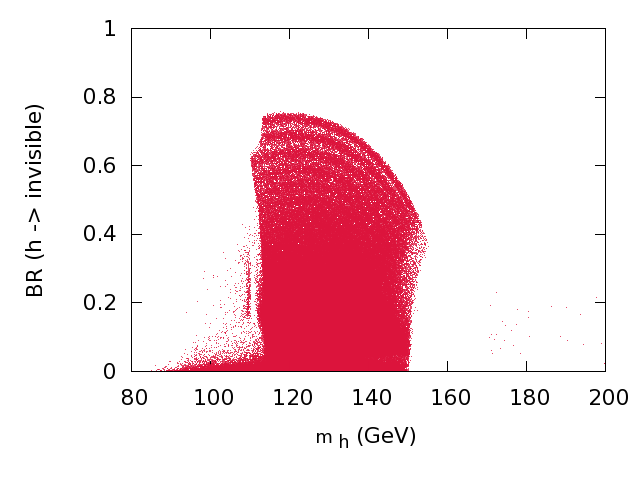}
\begin{tabular}{cc}
\includegraphics[scale=0.3,trim=0 0 0 0,clip=true]{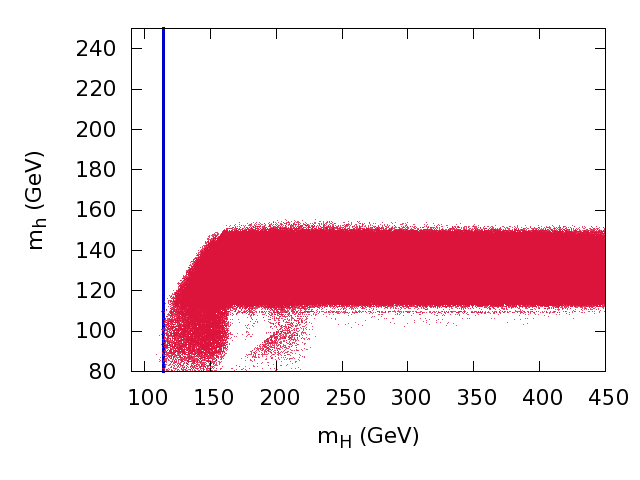}
&
\includegraphics[scale=0.3,trim=0 0 0 0,clip=true]{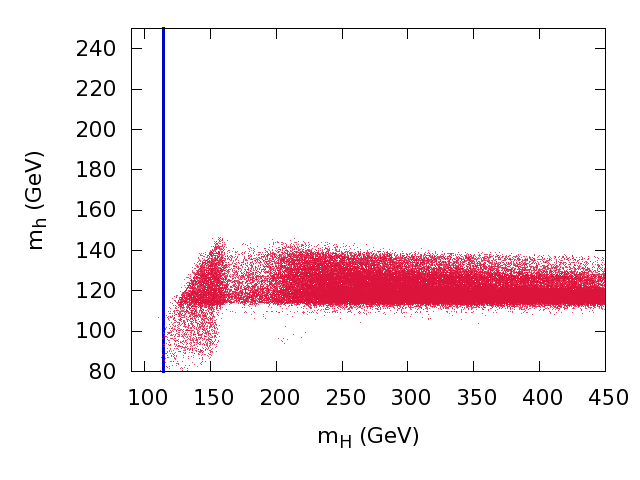}\\
\end{tabular}
\end{center}
\caption{\label{fig:BMSSM_inv} {\em Higgs decays to invisible neutralinos. The first graph shows the branching ratio of the
light Higgs to the lightest neutralinos, see text for details on the parameters of the neutralinos. In the second row, the first panel shows the allowed $m_h-m_H$ space taking into account
the present LHC constraint ($2$fb$^{-1}$). The graph on the right is for a luminosity of $15$fb$^{-1}$.}}
\end{figure}

Fig~\ref{fig:BMSSM_inv} shows the branching ratio of the light
Higgs to the lightest neutralinos. Between $m_h=120$ GeV and
$m_h=150$ GeV, the branching ratio is substantial ranging from
$~80\%$ to $40\%$ for $m_h=150$ GeV, at which point it drops
precipitously to almost $0\%$ because of the opening of the $WW$
channel. When the branching  into the lightest neutralinos is
large it reduces all the usual branchings and leads to a much
reduced sensitivity of the Higgs signal. Fig.~\ref{fig:BMSSM_inv}
shows how the picture changes when decays to invisibles are
allowed. With the current data  ($2$fb$^{-1}$) it is difficult to
see that changes have occurred. This is not surprising since our
invisible scenario can only cut in the $m_h$ range $120-150$ GeV .
With the present luminosity this range is still very much viable
even without Higgs decays as we have seen.  With the luminosity at
$15$fb$^{-1}$, we clearly see the {\em damaging} effect of the
invisible decays. More models with Higgs masses up to $m_h=140$GeV
survive compared to the case where no invisible Higgs decays are
allowed.

\subsection{Update from the preliminary results of the LHC with 5 fb$^{-1}$}
\begin{figure}[h!]
\begin{center}
\begin{tabular}{cc}
\includegraphics[scale=0.3,trim=0 0 0 0,clip=true]{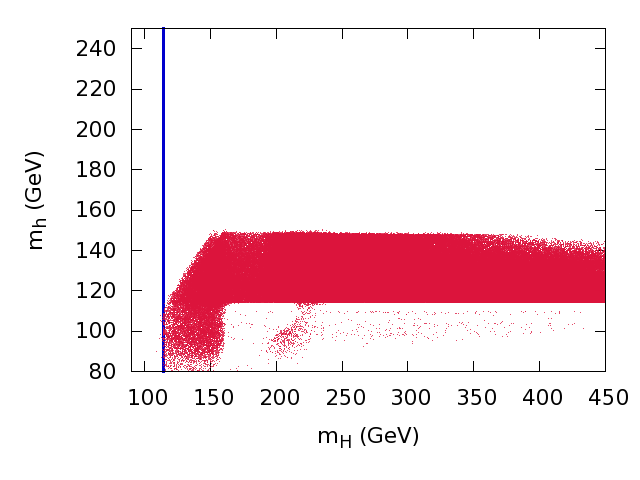}&
\includegraphics[scale=0.3,trim=0 0 0 0,clip=true]{BMSSM_1_5fb.png}\\
\end{tabular}
\end{center}
\caption{\label{fig:new} {\em Allowed range for $m_h$ and $m_H$ taking into account the  updated ATLAS and CMS  data with  $4.6-4.9$ fb$^{-1}$(left) compared to our previous analysis where a luminosity  rescaling to $5$ fb$^{-1}$ is applied to all channels(right), refer to the text for more details.}}
\end{figure}

A week after our paper was made public, the ATLAS and CMS
collaborations released the updates of some of their Higgs
analyses with $4.6$fb$^{-1}$ to $4.9$fb$^{-1}$ (see
\cite{atlas_5fb_comb,cms_5fb_comb} for combinations of the
different channels in each experiment). Since these results are
now available, we could update our analysis. The released data
point to an excess in some channels for $m_h \sim 124-126$GeV in
particular in the case of ATLAS where taking into account the {\em
look elsewhere effect} the global significance of the effect  is
$2.4$ standard deviations, thus not enough to claim any discovery.
In this section we therefore only consider the no signal
interpretation  and apply the 95\% exclusion limit for those
channels that have been reanalysed by ATLAS and CMS.  Compared to
the full listing given in section~\ref{subsub-incl} the update
includes i) for CMS: $\gamma\gamma$, $V\oo{b}b$, $\oo{\tau}\tau$,
$WW \to l\nu l\nu$, $ZZ \to 4l$ ii) for ATLAS: $\gamma\gamma$,
$ZZ\to 4l$, $WW\rightarrow l\nu l\nu$. We have kept the same data
set corresponding to $1$-$2.1$fb$^{-1}$ luminosity for the rest of
the list (with the ``inclusive" analysis). This is to be compared
to our projection for $5$fb$^{-1}$ where we applied a luminosity
rescaling on all the channels of the list. Our analysis of the
update is shown in fig. \ref{fig:new} in the plane $m_h-m_H$ . To
guide the eye and make the comparison easy, the figure also shows
the results based on the rescaling of all the channels. We see
little difference, the analysis based on a full rescaling of all
channels is, naturally, slightly more restrictive.

\section{Conclusions}

It is quite clear by now that Higgses in supersymmetric models
that go beyond the MSSM are very much constrained by the LHC
searches, even though the primary goal of those searches is the
Standard Model Higgs. We have first shown that the higher-order
terms appearing in the effective Lagrangian  alter  the Higgs
phenomenology quite significantly, in particular by raising the
lightest Higgs mass to values up to 250 GeV. This feature alone
was the main motivation of the BMSSM. We have shown that with the
advent of the LHC and the data collected so far, experiments no
longer allow a lightest supersymmetric Higgs to have a mass beyond
$150$ GeV even in these BMSSM set ups and even if we allowed for
decays into invisibles as provided by the lightest neutralinos.
With the increase of the luminosity most of the remaining models
at $15$fb$^{-1}$ are within a thin layer in lightest Higgs mass,
with $114<m_h<140$ GeV with a concentration around $m_h\sim
120$GeV, apart from an island with $m_h<100$ GeV for $m_H$ low
enough, $m_H<150$GeV. Invisible decays  allow more models with
$m_h \sim 140$GeV. Within this picture, set in terms of
exclusions, and with $15$fb$^{-1}$ of data, a similar conclusion
in terms of masses applies to the MSSM, the BMSSM lightest Higgs
is allowed to be less than about $10$GeV heavier that what it can
be in the MSSM, whereas before the advent of the LHC masses for
the lightest BSSM Higgs up to $250$GeV were possible.  Still the
phenomenology of the two models is quite different. Although our
philosophy in this paper has been towards constraining the BMSSM
models in the pessimistic prospect of no Higgs signal, it would be
very interesting to revisit the models in case of a signal. If the
density of allowed models that we have found is any indication for
where a possible signal may be hiding and if the possible slight
excess in the latest data from the LHC is confirmed,  it would be
extremely interesting to check whether the signals are better
described by a BMSSM Higgs with $m_h=125$ GeV and what the
properties of the latter are. Could one always tell it apart from
a MSSM one or even a standard model one? We have not addressed
this issue here. What we have addressed however, though perhaps
partially, is how to exploit LHC results made for the SM Higgs in
the context of other models that can have quite different
properties. We have made a request that the collaborations should
provide more details about the weight of the different
sub-channels that are used in their analyses.

\vspace*{1cm}
{\bf \large Acknowledgments}\\
We would like to thank Genevi\`eve B\'elanger for interesting
discussions throughout the study, Filip Moorgat for providing us
more insight on the CMS analyses. We also wish to thank Nicholas
Berger, Marta Felcini, Dmytro Kovalskyi, Biswarup Mukhopadhyaya,
Sasha Nikitenko, as well as the participants in the Higgs working
group at the workshop  {\em Physics at TeV Colliders} in Les Houches in June 2011.
Finally we acknowledge the exchanges with Marcela Carena
clarifying some points concerning the analyses in
Ref.\cite{carena_bmssmhiggs_1111}. This work is part of the French
ANR project, ToolsDMColl BLAN07-2 194882 and is supported in part
by the GDRI-ACPP of the CNRS (France).

\renewcommand{\thesection}{\Alph{section}}
\setcounter{section}{0}

\renewcommand{\theequation}{\thesection.\arabic{equation}}
\setcounter{equation}{0}

\newpage

\section{Appendix}
\subsection{Field-component Lagrangian}
\begin{equation}
\begin{split}
\mathcal{L}_K\ = \ &
    -(\oo{\textbf{W}}^{\oo{i}}+\1 K_{\overline{i}kl}\psi^k\psi^l)K_{\oo{i}j}(\textbf{W}^j+\1 K_{j\oo{k}\oo{l}}\psi^{\oo{k}}\psi^{\oo{l}})-\1|K_n\;gT^n|^2\textbf{W}_W^{\,-1}\hspace{1cm}\textrm{scalar potential}\\
    & +iK_{in}\;\partial_\mu\phi^i\;(gv)^{n\ \mu} \hspace{1cm}\textrm{gauge-scalars interaction}\\
    & +K_{nm}\;(gv)^{n\ \mu}\;(gv)^m_{\ \mu}+K_n\;(gv)^{n\ 2}\hspace{1cm}\textrm{gauge-gauge-scalars interaction}\\
    & +\1 K_{i\oo{j}}\;\partial_\mu\phi^i\;\partial^\mu\oo{\phi}^{\oo{j}}\hspace{1cm}\textrm{scalars derivatives}\\
    & +i\frac{\sigma^\mu}{2}K_{i\oo{j}}\;\partial_\mu\psi^i\;\oo{\psi}^{\oo{j}}+\1 \textbf{W}_{ij}\;\psi^i\;\psi^j+i\frac{\sigma^\mu}{2}K_{ij\oo{k}}\;\partial_\mu\phi^i\;\psi^j\;\oo{\psi}^{\oo{k}}\hspace{1cm}\textrm{fermion-fermion-scalars}\\
    & +i\sqrt{2}K_{in}\;\psi^i\;(g\lambda)^n+\sigma^\mu K_{i\oo{j}n}\;\psi^i\;\oo{\psi}^{\oo{j}}\;(gv)^n_{\ \mu}\hspace{1cm}\textrm{gauge-fermion-scalars interaction}\\
    & +\frac{1}{4}K_{ij\oo{k}\oo{l}}\;\psi^i\;\psi^j\;\oo{\psi}^{\oo{k}}\;\oo{\psi}^{\oo{l}}\hspace{1cm}\textrm{4 fermions-scalars interactions}\\
    & + \textbf{W}_W\,\left(-\frac{1}{4}F_{\mu\nu}F^{\mu\nu}-i\lambda\sigma^{\mu}D_{\mu}\oo{\lambda}\right)\hspace{1cm}\textrm{gauge interactions}\\
    & +\mathcal{L}_{SSB}\\
    & + h.c.i.n.
\end{split}\label{full_lag}
\end{equation}

\subsection{Kinetic rescaling}
Because of the kinetic Lagrangian for Higgs fields is non standard
\begin{equation}\LL_{\textrm{kinematics}}=d\phi^\dag K_2d\phi+i\oo{\psi}K_2\dslash\,\psi\end{equation}
we have to rescale the Higgs field in the following way.
$$\binom{h_1'}{h_2'}=\PD\binom{h_1}{h_2}\qquad\binom{\tilde{h}_1'}{\tilde{h}_2'}=\PD\binom{\tilde{h}_1}{\tilde{h}_2}$$
with $\PD=\sqrt{K_{i\oo{j}}}$. This quantity is straightforward to obtain from the effective K\"ahler potential \ref{effK}, and is the following, in terms of physical inputs
$$\PD=1-\frac{1}{1+\tb^2}\frac{v_0^2}{M^2}
\begin{pmatrix}
4a_{10}+4\tb a_{50}+(a_{30}+a_{40})\tb^2 & a_{50}+\tb(a_{30}+a_{40})+\tb^2 a_{60}\\
a_{50}+\tb(a_{30}+a_{40})+\tb^2 a_{60} & a_{30}+a_{40}+4\tb a_{60}+4\tb^2 a_{20}
\end{pmatrix}
\label{PD}
$$
$v_0^2$ is defined solely in terms of the Standard Model physical inputs $(e,M_W,M_Z$), that is $v_0^2=2\frac{M_Z^2s_W^2c_W^2}{e^2}$

\newpage


\end{document}